\documentclass{aastex}
\usepackage{emulateapj5,times,mathptm}

\newcommand {\asec} {$^{\prime\prime}$}

\def\amin{\ifmmode ^{\prime}\else$^{\prime}$\fi}
\def\asec{\ifmmode ^{\prime\prime}\else$^{\prime\prime}$\fi}

\def\etal{{et\,al.\,}}

\def\asca{{\it ASCA\/}}

\def\chandra{{\it Chandra\/}}

\def\heao1{{\it HEAO-1\/}}

\def\xmm{{\it XMM-Newton\/}}

\def\ltsima{$\; \buildrel < \over \sim \;$}
\def\simlt{\lower.5ex\hbox{\ltsima}}
\def\gtsima{$\; \buildrel > \over \sim \;$}
\def\simgt{\lower.5ex\hbox{\gtsima}}

\journalinfo{The Astronomical Journal, 2001 November, astro-ph/0107450}

\begin{document}
%%%%%%%%%%%%%%%%%%%%%%%%%%%%%%%%%%%%%%%%%%%%%%%%%%%%%%%%%%%%%%%%%%%%%%%%%%%%%%%%%%
%

%
%%%%%%%%%%%%%%%%%%%%%%%%%%%%%%%%%%%%%%%%%%%%%%%%%%%%%%%%%%%%%%%%%%%%%%%%%%%%%%%%%%
\title{The Chandra Deep Field North Survey. VI. The Nature of the Optically Faint X-ray Source Population}
%%%%%%%%%%%%%%%%%%%%%%%%%%%%%%%%%%%%%%%%%%%%%%%%%%%%%%%%%%%%%%%%%%%%%%%%%%%%%%%%%%
%

\author{D.M.~Alexander,$^1$ W.N.~Brandt,$^1$ A.E.~Hornschemeier,$^1$ G.P.~Garmire,$^1$ D.P.~Schneider,$^1$ F.E.~Bauer$^1$ and R.E.~Griffiths$^2$}

\footnotetext[1]{Department of Astronomy \& Astrophysics, 525 Davey Laboratory, The Pennsylvania State University, University Park, PA 16802}

\footnotetext[2]{Department of Physics, Carnegie Mellon University, Pittsburgh, PA 15213}

%\vfill\eject

\shorttitle{OPTICALLY FAINT X-RAY SOURCES IN THE CHANDRA DEEP FIELD NORTH SURVEY}

\shortauthors{ALEXANDER ET AL.}

\slugcomment{Received 2001 May 14; accepted 2001 July 19}

%
%%%%%%%%%%%%%%%%%%%%%%%%%%%%%%%%%%%%%%%%%%%%%%%%%%%%%%%%%%%%%%%%%%%%%%
\begin{abstract}
%%%%%%%%%%%%%%%%%%%%%%%%%%%%%%%%%%%%%%%%%%%%%%%%%%%%%%%%%%%%%%%%%%%%%%
%

We provide constraints on the nature of the optically faint ($I\ge24$) X-ray source population from a 1~Ms \chandra\ exposure of a $8.4\amin\times8.4\amin$ region within the Hawaii flanking-field area containing the Hubble Deep Field North region. We detect 47 ($2,400^{+400}_{-350}$ deg$^{-2}$) optically faint sources down to 0.5--2.0~keV and 2.0--8.0~keV fluxes of $\approx~3\times10^{-17}$~erg~cm$^{-2}$~s$^{-1}$ and $\approx~2\times10^{-16}$~erg~cm$^{-2}$~s$^{-1}$, respectively; these sources contribute $\approx 14$\% and $\approx 21$\% of the 0.5--2.0~keV and 2.0--8.0~keV X-ray background radiation, respectively. The fraction of optically faint X-ray sources is approximately constant (at $\approx 35$\%) for 0.5--8.0~keV fluxes from $3\times10^{-14}$~ergs~cm$^{-2}$~s$^{-1}$ down to the X-ray flux limit. A considerable fraction (30$^{+14}_{-10}$\%) of the optically faint X-ray sources are Very Red Objects ($I-K\ge$~4). Analysis of the optical and X-ray properties suggests a large number of optically faint X-ray sources are likely to host obscured AGN activity at $z=$~1--3. From these results we calculate that a significant fraction ($\approx$~5--45\%) of the optically faint X-ray source population could be obscured QSOs (rest-frame unabsorbed 0.5--8.0~keV luminosity $>3\times10^{44}$~erg~s$^{-1}$) at $z\le3$. Given the number of X-ray sources without $I$-band counterparts, there are unlikely to be more than $\approx$~15 sources at $z>6$. We provide evidence that the true number of $z>6$ sources is considerably lower. 

We investigate the multi-wavelength properties of optically faint X-ray sources. Nine optically faint X-ray sources have $\mu$Jy radio counterparts; $\approx 53^{+24}_{-17}$\% of the optically faint $\mu$Jy radio sources in this region. The most likely origin of the X-ray emission in these X-ray detected, optically faint $\mu$Jy radio sources is obscured AGN activity. However, two of these sources have been previously detected at sub-millimeter wavelengths and the X-ray emission from these sources could be due to luminous star formation activity. Assuming the spectral energy distribution of NGC~6240, we estimate the 175~$\mu$m flux of a typical optically faint X-ray source to be \hbox{$<10$~mJy}; however those sources with detectable sub-millimeter counterparts (i.e.,\ $f_{850{\rm \mu m}}>$~3~mJy) could be substantially brighter. Hence, most optically faint X-ray sources are unlikely to contribute significantly to the far-IR (140--240~$\mu$m) background radiation. However, as expected for sources with AGN activity, the two optically faint X-ray sources within the most sensitive area of the {\it ISOCAM} HDF-N region have faint ($\simlt$50~$\mu$Jy) 15~$\mu$m counterparts. 

We also provide constraints on the average X-ray properties of classes of optically faint sources not individually detected at X-ray energies. Stacking analyses of optically faint $\mu$Jy radio sources not individually detected with X-ray emission yields a possible detection (at $98.3$\% confidence) in the 0.5--2.0~keV band; this X-ray emission could be produced by star formation activity at $z=$~1--3. None of the optically faint AGN-candidate sources in the HDF-N itself are detected at X-ray energies either individually or with stacking analyses, showing that these sources have low X-ray luminosities if they are indeed AGN.

\end{abstract}


%
%%%%%%%%%%%%%%%%%%%%%%%%%%%%%%%%%%%%%%%%%%%%%%%%%%%%%%%%%%%%%%%%%%%%%%
\section{Introduction}\label{intro}
%%%%%%%%%%%%%%%%%%%%%%%%%%%%%%%%%%%%%%%%%%%%%%%%%%%%%%%%%%%%%%%%%%%%%%
%

One of the key goals of X-ray astronomy during the last 40 years has been to determine the origin of the X-ray background (Giacconi \etal 1962). Surveys taken prior to the launch of the {\it Chandra X-ray Observatory} (hereafter \chandra; Weisskopf \etal 2000) performed in both the soft X-ray ($\approx$~0.5--2.0~keV) and hard X-ray ($\approx$~2.0--10.0~keV) bands showed that a significant fraction of the X-ray background is produced by discrete sources, primarily obscured and unobscured AGN (e.g., Hasinger \etal 1998; Ueda \etal 1998, 1999; Fiore \etal 1999; Akiyama \etal 2000; Lehmann \etal 2001). The improved sensitivity and sub-arcsecond resolution of {\it Chandra} is allowing deep X-ray surveys to resolve close to 100\% of the $\approx$~0.5--8.0~keV background (e.g.,\ Mushotzky \etal 2000; Brandt \etal 2001a, hereafter Paper IV; Garmire \etal 2001, hereafter Paper III; Tozzi \etal 2001; Brandt \etal 2001b, hereafter Paper V). The optical spectroscopic identification of the optically brighter ($I<24$) X-ray sources is currently in progress, and the majority of the sources appear to be AGN at $z\simlt$1 (e.g., Hornschemeier \etal 2001, hereafter Paper II; Tozzi \etal 2001; A.J. Barger \etal, in preparation). However, a significant fraction ($\approx 30$\%) of the X-ray sources are too faint ($I\ge24$) for optical spectroscopic observations (e.g.,\ Barger \etal 2001a; Paper II; Tozzi \etal 2001). 

The combination of faint optical emission and bright X-ray emission suggests that many of these optically faint X-ray sources are powerful high-redshift AGN (e.g.,\ Fabian \etal 2000; Barger \etal 2001a; Cowie \etal 2001; Schreier \etal 2001). The nature and properties of such sources are important for understanding moderate-to-high-redshift ($1<z<7$) accretion activity and the role of AGN activity in galaxy formation (e.g.,\ Fabian 1999; Haiman \& Loeb 1999; Cowie \etal 2001). Some optically faint X-ray sources have flat X-ray spectral slopes (e.g.,\ Paper II; Cowie \etal 2001), suggesting they may be highly obscured AGN, and a number of the sources could be obscured QSOs (i.e.,\ $L_X>3\times10^{44}$ ergs$^{-1}$). Although very few obscured QSOs have been detected locally, they are predicted to exist in large numbers at high redshift (e.g.,\ Wilman, Fabian, \& Nulsen 2000; Gilli, Salvati, \& Hasinger 2001). If the origin of the obscuration in these sources are optically thick dusty tori (e.g.,\ Antonucci 1993), they should also produce powerful infrared emission and may contribute significantly to the cosmic infrared background (e.g.,\ Puget \etal 1996; Schlegel \etal 1998).

Significant numbers of optically faint X-ray sources have been detected in other X-ray surveys; however, the analysis of such sources has been either limited to detailed single object studies (e.g.,\ Cowie \etal 2001) or combined within larger object identification studies (e.g.,\ Mushotzky \etal 2000; Barger \etal 2001a; Paper II). We present here the first detailed analysis of the optically faint ($I\ge24$) X-ray source population with a 1~Ms \chandra\ observation of the HDF-N and surrounding Hawaii flanking-field area (i.e.,\ Paper V).\footnote{The Hawaii flanking-field area is defined by the optical and near-IR observations presented in Barger \etal (1999).} We have chosen the Hawaii flanking-field area in this study for a number of important reasons. First, we include the most sensitive and positionally accurate X-ray data as the HDF-N itself is at the aim-point of the \chandra\ observation. Second, the Hawaii flanking-field area has deep optical (Barger \etal 1999), radio (Richards \etal 1998; Richards 2000) and sub-millimeter (Barger, Cowie, \& Richards 2000; Chapman \etal 2001; Barger \etal 2001b) coverage, and the properties of optically faint $\mu$Jy radio sources within this region have been pursued by Richards \etal (1999). Finally, by including the HDF-N itself we guarantee very deep multi-wavelength coverage within the central region. The high surface density of $I\ge24$ sources means that \chandra\ is the only X-ray observatory with the positional accuracy to pin-point optically faint X-ray sources. The larger positional uncertainty of \xmm\ detected sources will sometimes result in $>1$ optically faint counterpart to an X-ray source.

In this study we compare the X-ray, optical and near-IR properties of the optically faint X-ray sources to those of the optically bright X-ray sources. We investigate the radio and infrared properties of optically faint X-ray sources, review the properties of the best-studied optically faint X-ray sources to date, estimate their redshifts and place constraints on the fraction of obscured QSOs in the optically faint X-ray source population. We also provide constraints on the X-ray emission properties of optically faint $\mu$Jy radio sources (Richards \etal 1999) not individually detected at X-ray energies and optically faint AGN candidates (Jarvis \& MacAlpine 1998; Conti \etal 1999) within the HDF-N itself.

The Galactic column density along this line of sight is $(1.6\pm 0.4)\times 10^{20}$~cm$^{-2}$ (Stark \etal 1992), and $H_0=70$~km~s$^{-1}$ Mpc$^{-1}$ and $q_0=0.1$ are adopted throughout this paper. All coordinates in this paper are J2000.

%
%%%%%%%%%%%%%%%%%%%%%%%%%%%%%%%%%%%%%%%%%%%%%%%%%%%%%%%%%%%%%%%%%%%%%%
\section{Chandra ACIS-I observations}
%%%%%%%%%%%%%%%%%%%%%%%%%%%%%%%%%%%%%%%%%%%%%%%%%%%%%%%%%%%%%%%%%%%%%%
%

The X-ray results reported in this paper were obtained with the 1~Ms \chandra\ Advanced CCD Imaging Spectrometer (ACIS; G.P. Garmire \etal, in preparation) survey of the \hbox{HDF-N} and its environs presented in Paper V. With the exception of a number of lower significance \chandra\ sources reported in \S5.1, all the \chandra\ sources were taken from \hbox{Paper V}. Results for the HDF-N itself have been presented by Hornschemeier \etal (2000, hereafter Paper I) and in Paper IV for 164.5~ks and 479.7~ks exposures, respectively. Results obtained with a 221.9~ks exposure over the larger $8.6\amin\times8.7\amin$ Caltech Faint Field Galaxy Redshift Survey Area (hereafter referred to as the ``Caltech area''; e.g.,\ Cohen \etal 2000; Hogg \etal 2000) centered on the HDF-N have been presented in Paper II. The area used in this study ($8.4\amin\times8.4\amin$) is slightly smaller than the Hawaii Flanking Field area, and a large fraction of it overlaps with the Caltech area.\footnote{We have not used the full Hawaii flanking field area to avoid the lower optical and near-IR sensivity towards the edge of these images.} Source detection in each standard X-ray band (see Paper V) was performed with {\sc wavdetect} (Freeman \etal 2001) with a probability threshold of 10$^{-7}$; we would expect $\approx 0.11$ spurious sources for each X-ray band over the entire region with this procedure. A lack of spurious faint X-ray sources is essential for this study since some sources may not have a detectable optical counterpart; we provide corroborating evidence for a low fraction of spurious optically faint X-ray sources in \S3.2. All sources were inspected to ensure that they are not produced or affected by ``cosmic ray afterglows'' (\chandra\ X-ray Center 2000, private communication). 

In total 141 sources (hereafter referred to as the entire X-ray sample) were detected in the $8.4\amin\times8.4\amin$ region defined here: 136 in the full band (0.5--8.0~keV), 117 in the soft band (0.5--2.0~keV) and 102 in the hard band (2.0--8.0~keV). We show the \chandra\ image in Figure~1. The ``effective'' full-band exposure time per source, as derived from our exposure map, ranges from 644--945~ks with most sources (85\%) having $>800$~ks of exposure. Even with these long exposure times, the \chandra\ ACIS is entirely photon limited for point-source detection near the aim point. For a power-law model with photon index $\Gamma=1.4$ and the Galactic column density, our $\approx 6$ count soft-band and $\approx 10$ count hard-band detection limits correspond to flux limits of \hbox{$\approx 3\times 10^{-17}$~erg~cm$^{-2}$~s$^{-1}$} and \hbox{$\approx 2\times 10^{-16}$~erg~cm$^{-2}$~s$^{-1}$}, respectively. The absolute X-ray source positions within 5\amin\ of the aim point are accurate to 0.6\asec; for sources outside this region, the positional errors rise to $\approx 1$\asec (see Paper V).

\begin{figure*}
\vspace{0.0in}
\centerline{\includegraphics[width=18.0cm]{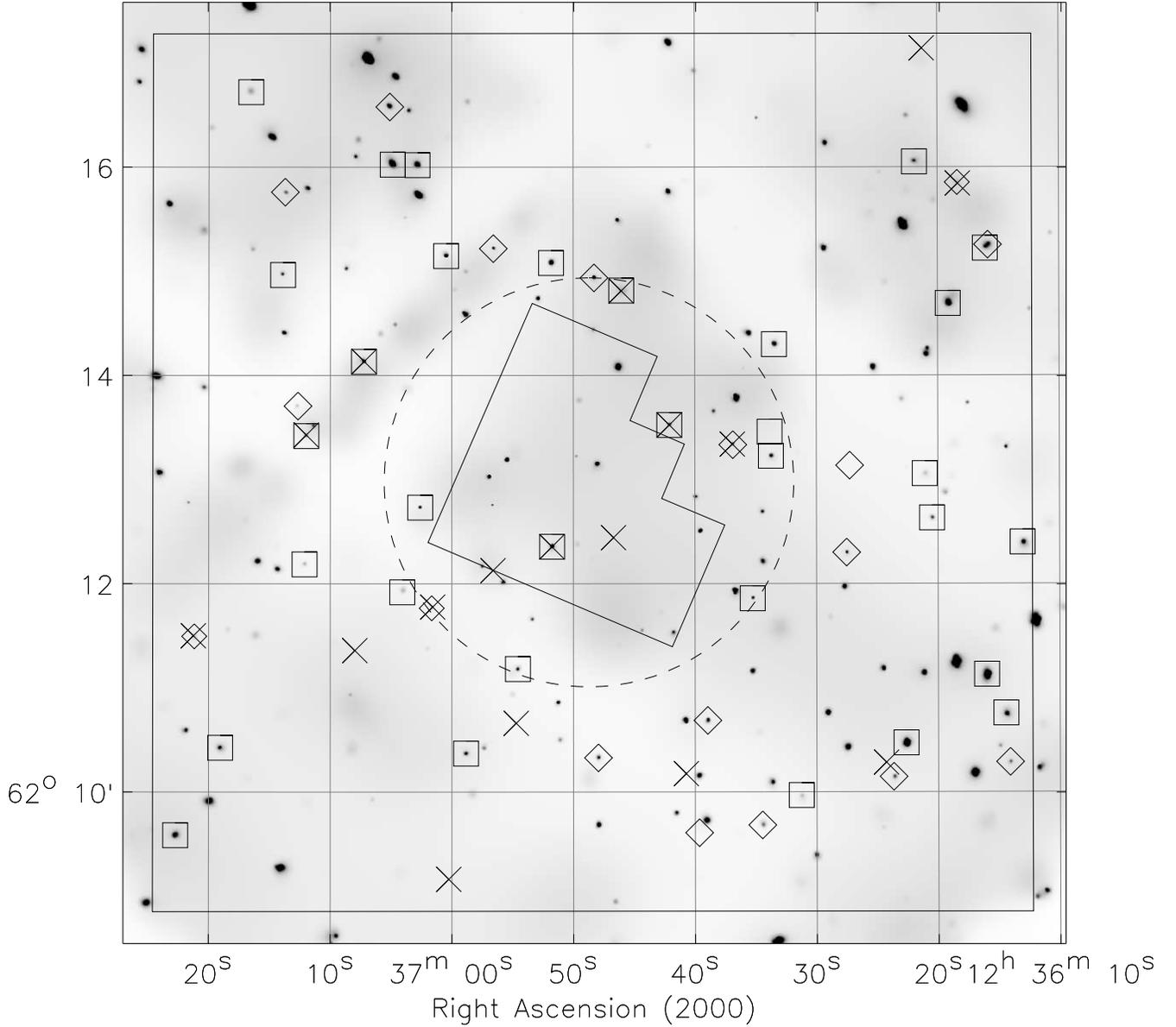}}
\vspace{0.5in}
\figcaption{Adaptively smoothed full-band \chandra\ image. The squares show the positions of the optically faint X-ray sources with 2$\sigma$ $I$-band counterparts, the diamonds show the positions of the optically faint X-ray sources without 2$\sigma$ $I$-band counterparts, and the crosses show the positions of the 17 optically faint $\mu$Jy radio sources from Richards \etal (1999). This image has been made using the standard \asca\ grade set and has been adaptively smoothed at the $3\sigma$ level using the code of Ebeling \etal (2001). The HDF-N is shown as the polygon at the center of the image, the large box indicates the $8.4\amin\times8.4\amin$ region used in this study, and the dashed circle indicates the 2\amin\ radius from the center of the HDF-N; this corresponds to the approximate region covered by the 15~$\mu$m {\it ISOCAM} survey (Serjeant \etal 1997; Aussel \etal 1999). Most of the apparent diffuse emission is instrumental background. Clearly some of the optically faint sources are among the brightest X-ray sources in the entire X-ray sample. The faintest X-ray sources are below the significance level of the smoothing and thus are not visible in this figure.}
\label{sourceexp}
\end{figure*}

%%%%%%%%%%%%%%%%%%%%%%%%%%%%%%%%%%%%%%%%%%%%%%%%%%%%%%%%%%%%%%%%%%%%%%

%
%%%%%%%%%%%%%%%%%%%%%%%%%%%%%%%%%%%%%%%%%%%%%%%%%%%%%%%%%%%%%%%%%%%%%%
\section{Basic optical and X-ray properties of the optically faint X-ray sources}
%%%%%%%%%%%%%%%%%%%%%%%%%%%%%%%%%%%%%%%%%%%%%%%%%%%%%%%%%%%%%%%%%%%%%%
%

\subsection{Optical source magnitudes and spectroscopic identifications}

There are published $\approx 5\sigma$ magnitudes for $I<24.3$ and $HK^\prime<20.4$ sources over the entire region used in this study (Barger \etal 1999). We retrieved the publicly available images\footnote{These images are available at http://www.ifa.hawaii.edu/$\sim$cowie/hdflank/hdflank.html.} to search for fainter sources and determine magnitudes for X-ray sources down to $\approx 2\sigma$ limits of $I=25.3$ and $HK^\prime=21.4$ using the {\sc sextractor} photometry tool (Bertin \& Arnouts 1996), assuming the ``Best'' magnitude criteria; see Table 1. We found good agreement ($\approx 1\sigma$ magnitude deviations of $\pm$0.25 mags) with Barger \etal (1999) for the sources included in their catalog. While a large fraction of this region has additional optical-to-near-IR coverage (Hogg \etal 2000), we have not included analysis of this data since deeper multi-band optical photometry is currently being obtained (A.J. Barger \etal, in preparation). Although we only have two optical-to-near-IR magnitudes per source, the choice of $I$-band and $HK^\prime$-band observations are useful as many sources have red optical-to-near-IR colors (see \S4.1). The addition of deep shorter wavelength 
%
%%%%%%%%%%%%%%%%%%%%%%%%%%%%%%%%%%%%%%%%%%%%%%%%%%%%%%%%%%%%%%%%%%%%%%
% 2 I-band magnitude distribution
%%%%%%%%%%%%%%%%%%%%%%%%%%%%%%%%%%%%%%%%%%%%%%%%%%%%%%%%%%%%%%%%%%%%%%
%
\vspace{0.2in}
\centerline{\includegraphics[angle=-90,width=9.0cm]{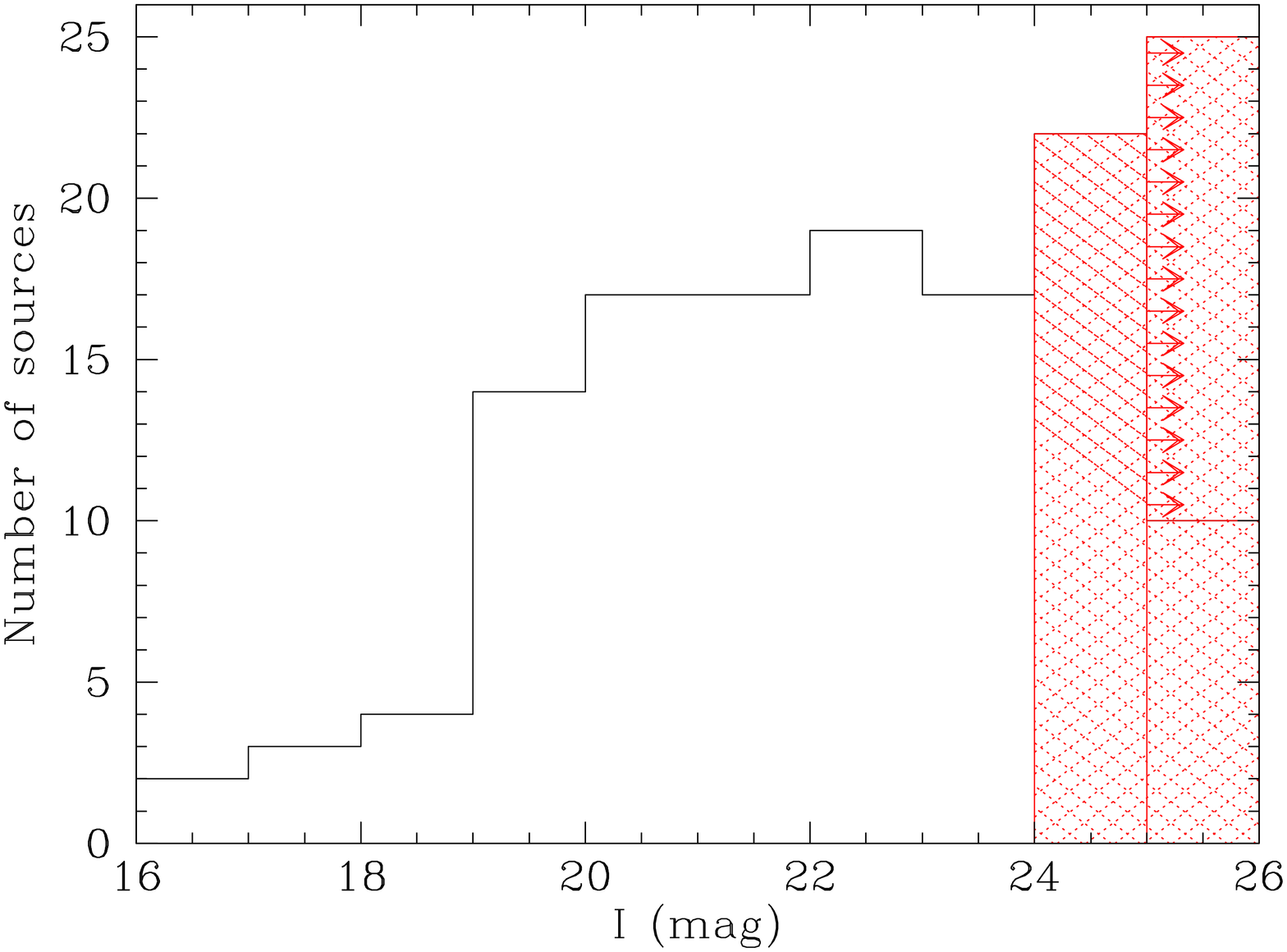}}
\vspace{0.1in}
\figcaption{The $I$-band magnitude distribution of optical counterparts for the entire X-ray sample. The hatched blocks are the optically faint sources, and the unhatched blocks are the optically bright sources; the 15 sources without 2$\sigma$ $I$-band counterparts are shown as upper limit arrows which correspond to $I>25.3$. Compare to Figure~1 of Richards \etal (1999).}
\label{fig:imag}
\vspace{0.2in}

%%%%%%%%%%%%%%%%%%%%%%%%%%%%%%%%%%%%%%%%%%%%%%%%%%%%%%%%%%%%%%%%%%%%%%

\noindent optical observations would also be useful and could provide strong constraints on source redshifts through the utilization of the band drop-out technique (e.g.,\ Steidel \etal 1996; see \S6.2).

We matched the X-ray sources to $I$-band counterparts using a search radius of 1\asec; the $I$-band magnitude distribution of the entire X-ray sample is shown in Figure~2. This distribution can be compared to the $\mu$Jy radio source $I$-band magnitude distribution of Richards \etal (1999), whose survey covered approximately the same region. The $I$-band magnitude distribution of the radio sources peaks with a median of $I\approx 22$ and falls off at fainter magnitudes. At $I>25$ a population of optically faint $\mu$Jy radio sources is detected; 30\% of the $\mu$Jy radio sources have $I>25$. The median $I$-band magnitude of the entire X-ray sample is also $I\approx 22$, although we do not see a decline to fainter optical magnitudes; in fact the distribution is reasonably flat for $I>20$ (16\% have $I>25$).

We have adopted $I\ge24$ as the definition of an optically faint source. This optical magnitude limit is fainter than that which can be reasonably achieved with optical spectroscopy on a 10-m class telescope, and therefore other techniques are required to determine the redshifts and nature of these sources. From the entire X-ray sample, 47 sources have $I\ge24$ (see Table 1) and 22 of these have an $HK^\prime$-band counterpart. These sources account for $33^{+6}_{-5}$\% of the X-ray sources detected in the entire X-ray sample.\footnote{Note that all errors are determined from Tables 1 and 2 of Gehrels (1986) and correspond to the $1\sigma$ level.} Thumbnail images of all the optically faint X-ray sources are shown in Figure~3. None of the sources show evidence for extended emission (although also see \S4.5); the slightly extended structure of CXOHDFN J123619.2+621442 is due to a nearby faint X-ray source. However, one source (CXOHDFN J123616.1+621514) appears to be associated with an optically blank X-ray source (see \S6.2 for further discussion). Only one source (CXOHDFN J123651.8+621221; see Paper I and Paper IV) lies in the HDF-N itself; this source has an $F814W=25.8$ counterpart (M.~Dickinson 2000, private communication).

%
%%%%%%%%%%%%%%%%%%%%%%%%%%%%%%%%%%%%%%%%%%%%%%%%%%%%%%%%%%%%%%%%%%%%%%
% 3 Thumbnail images of all the X-ray sources
%%%%%%%%%%%%%%%%%%%%%%%%%%%%%%%%%%%%%%%%%%%%%%%%%%%%%%%%%%%%%%%%%%%%%%
%

\begin{figure*}
\vspace{0.0in}
\centerline{\includegraphics[angle=0,width=23.0cm]{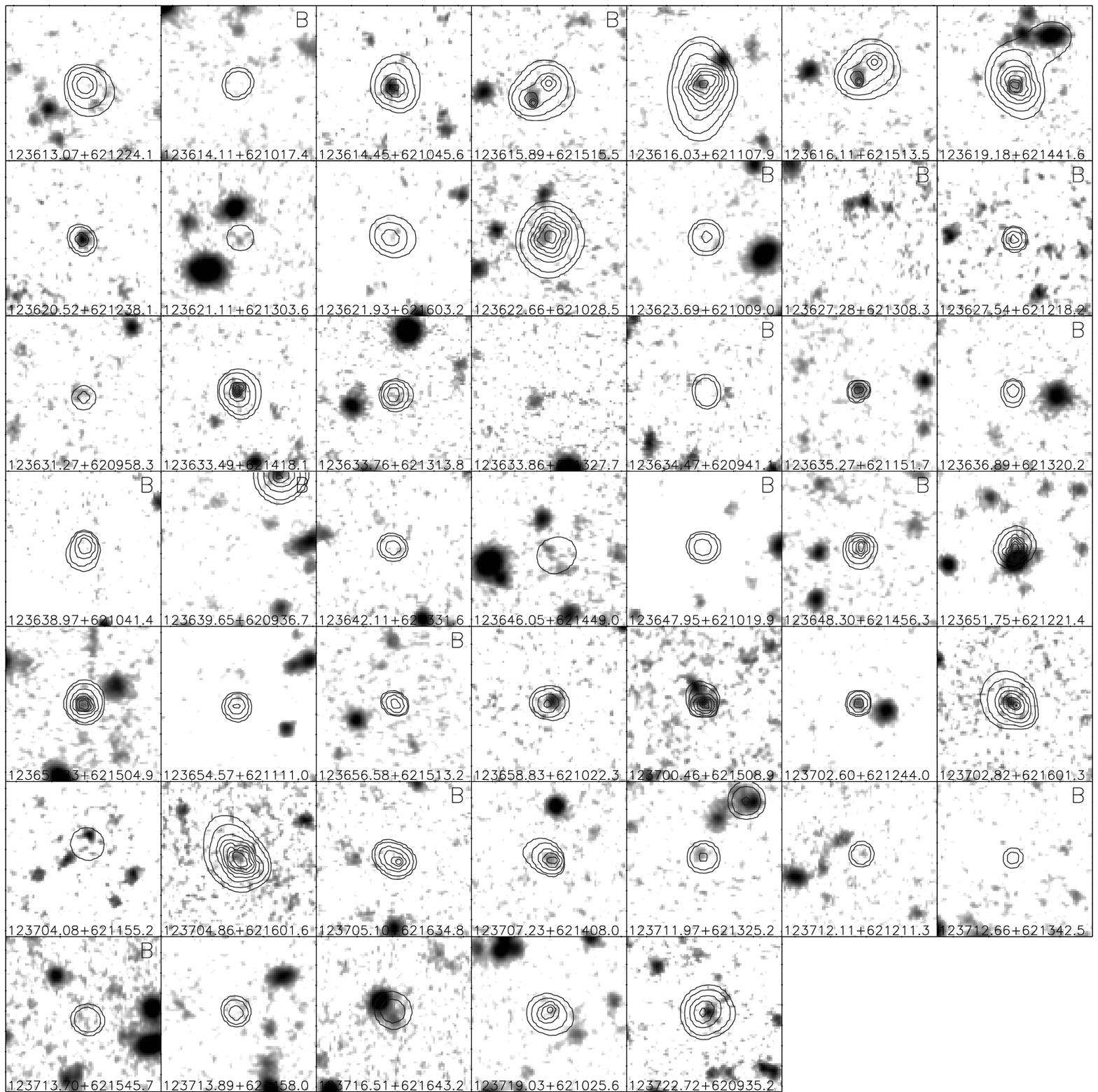}}
\vspace{0.5in}
\figcaption{Thumbnail $I$-band images with overlaid X-ray contours for each source in the optically faint X-ray sample. The contour levels refer to the number of counts detected in the full band; each X-ray image has been adaptively smoothed and is $15.09\asec$ on a side. It is clear from these images that a number of X-ray sources have no visible 2$\sigma$ $I$-band counterpart; the optically blank sources listed in Table~1 are indicated here with a ``B''. The faintest X-ray sources are below the significance level of the smoothing and thus are not visible in these figures. One optically faint X-ray source (CXOHDFN J123616.1+621514) appears to be interacting with an optically blank X-ray source.}
\label{sourceexp}
\end{figure*}

%%%%%%%%%%%%%%%%%%%%%%%%%%%%%%%%%%%%%%%%%%%%%%%%%%%%%%%%%%%%%%%%%%%%%%

Fifteen sources do not have 2$\sigma$ $I$-band counterparts, setting an $I$-band magnitude limit of $I>25.3$, although four of these sources have $HK^\prime$ counterparts. These sources are referred to as optically blank X-ray sources and remain a part of the optically faint X-ray source sample, although we test whether they are statistically different in \S6.2. Given the surface density of $I<24$ sources there is an $\approx 2.3$\% chance that an X-ray source is coincidently matched to a $I<24$ source. Therefore we would expect $\approx 2.2$ optically faint X-ray sources to have been erroneously matched to an optically bright counterpart; as this is a small fraction of the optically faint X-ray source sample, this discrepancy will not affect our conclusions.

Published optical spectroscopic identifications are presently available for 61 (65\%) of the optically bright ($I<24$) X-ray sources; these spectroscopic identifications were taken from Cohen \etal (2000), Paper II, and Dawson \etal (2001). For comparison, only two optically faint X-ray sources have spectroscopic identifications. One source (CXOHDFN J123633.5+621418) is a $z=3.403$ (see Cohen \etal 2000; Paper II) broad-line AGN (BLAGN) that is only marginally optically faint with our photometry ($I=24.1$). The other source (CXOHDFN J123642.1+621332) is fainter and lies at a higher redshift ($z=4.424$; Waddington \etal 1999; but also see \S2.2 of Barger, Cowie, \& Richards 2000). The spectroscopic and X-ray properties of this source strongly suggest it contains an AGN (Waddington \etal 1999; Paper IV).

To assist in the interpretation of the optically faint X-ray sample we have defined two sub-samples from the optically bright X-ray source sample. The first includes seven of the eight BLAGN reported in Paper II; the eighth source is optically faint with our photometry (see above). The second contains luminous narrow-line AGN (NLAGN) and includes the six narrow-line X-ray sources with 0.5--8.0~keV luminosities $>3\times10^{42}$ erg~s$^{-1}$ reported in Paper II. Three of the narrow-lined sources have AGN signatures in their optical spectra, while the signal-to-noise in the optical spectroscopic observations of the other three sources is insufficient to detect high ionization AGN lines (although their high X-ray luminosities suggest they are NLAGN; Paper II). While NLAGN sources with X-ray luminosities lower than $3\times10^{42}$ erg~s$^{-1}$ were also reported in Paper II, we have chosen this definition of a NLAGN to minimize the potential contribution at X-ray energies from starburst activity. These samples are not complete, as a fraction of the unclassified optically bright X-ray sources are likely to be BLAGN or luminous NLAGN.

\subsection{X-ray fluxes}

The full-band X-ray flux distribution of the entire X-ray sample is shown in Figure~4. The optically faint sources are detected over a similar X-ray flux range to the optically bright sources, and clearly some optically faint sources are among the brightest X-ray sources in the entire X-ray sample (see also Figure~1). Similar percentages of optically faint and optically bright X-ray sources are detected in the full band ($98^{+2}_{-14}$\% for the optically faint sample versus $96^{+4}_{-10}$\% for the optically bright sample). A Kolmogorov-Smirnov (K-S) test shows that the full-band X-ray flux distributions of these sources are indistinguishable; the K-S test probability is 45\%. This provides additional evidence that the number of spurious optically faint sources is low (see \S2). The fraction of optically faint and optically bright X-ray sources detected in the hard band ($83^{+16}_{-13}$\% for the
%
%%%%%%%%%%%%%%%%%%%%%%%%%%%%%%%%%%%%%%%%%%%%%%%%%%%%%%%%%%%%%%%%%%%%%%
% 4 X-ray fluxes
%%%%%%%%%%%%%%%%%%%%%%%%%%%%%%%%%%%%%%%%%%%%%%%%%%%%%%%%%%%%%%%%%%%%%%
%
\vspace{0.2in}
\centerline{\includegraphics[angle=0,width=9.0cm]{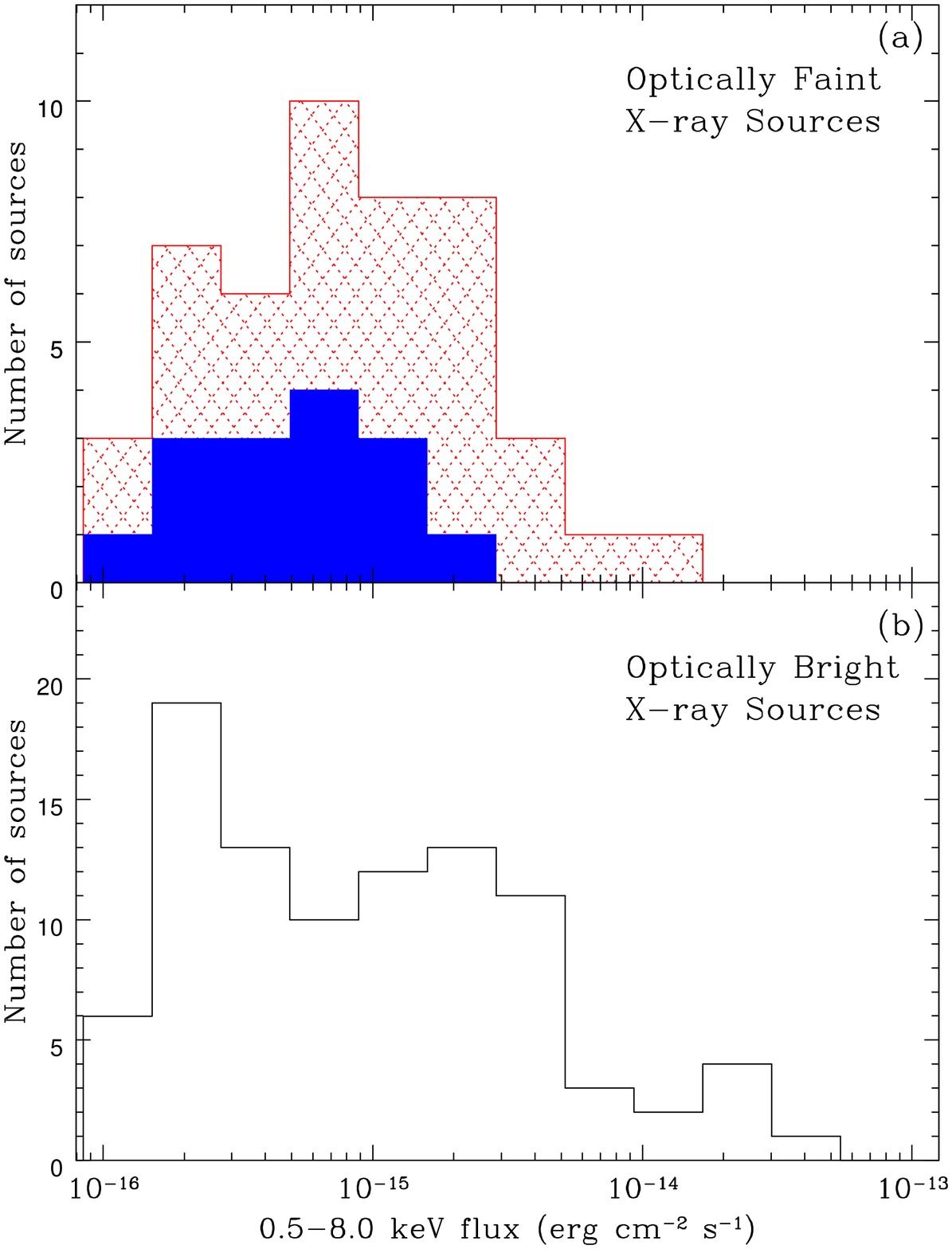}}
\figcaption{The full-band flux distributions of (a) the optically faint X-ray sample and (b) the optically bright X-ray sample. The two X-ray flux distributions are not distinguishable according to the Kolmogorov-Smirnov (K-S) test. The solid blocks show the overlaid flux distribution of X-ray sources without 2$\sigma$ $I$-band counterparts.}
\label{sourceexp}
\vspace{0.2in}

%%%%%%%%%%%%%%%%%%%%%%%%%%%%%%%%%%%%%%%%%%%%%%%%%%%%%%%%%%%%%%%%%%%%%%

\noindent optically faint sample versus $67^{+10}_{-8}$\% for the optically bright sample) and soft band ($74^{+15}_{-13}$\% for the optically faint sample versus $87^{+11}_{-10}$\% for the optically bright sample) are statistically consistent, although the current uncertainties are large.

In Figure~5 we show the fraction of optically faint sources in the entire X-ray sample versus full-band flux. With the exception of the brightest X-ray flux bin, where only one optically bright X-ray source is detected, the data are consistent with an $\approx 35$\% fraction of optically faint sources for full-band fluxes of $<3\times 10^{-14}$~erg~cm$^{-2}$~s$^{-1}$. We have not detected any optically faint sources with soft-band X-ray fluxes $>3\times 10^{-15}$~erg~cm$^{-2}$~s$^{-1}$ or hard-band fluxes $>2\times 10^{-14}$~erg~cm$^{-2}$~s$^{-1}$; however, optically faint X-ray sources with larger X-ray fluxes have been detected in other \chandra\ surveys (e.g.,\ source 7 in Mushotzky \etal 2000 has a soft-band flux of $1.5\times 10^{-14}$~erg~cm$^{-2}$~s$^{-1}$ and a hard-band X-ray flux of $3.8\times 10^{-14}$~erg~cm$^{-2}$~s$^{-1}$). Wide-field, shallow X-ray surveys such as the \chandra\ Multi-wavelength Project (ChaMP; Wilkes \etal 2001)\footnote{Details of the ChaMP project can be found at http://hea-www.harvard.edu/CHAMP/.} will be well suited for determining the fraction of optically faint sources at brighter X-ray fluxes (i.e.,\ full-band X-ray fluxes $>5\times 10^{-15}$~erg~cm$^{-2}$~s$^{-1}$). 

%
%%%%%%%%%%%%%%%%%%%%%%%%%%%%%%%%%%%%%%%%%%%%%%%%%%%%%%%%%%%%%%%%%%%%%%
% 5 X-ray flux fraction
%%%%%%%%%%%%%%%%%%%%%%%%%%%%%%%%%%%%%%%%%%%%%%%%%%%%%%%%%%%%%%%%%%%%%%
%

\vspace{0.2in}
\centerline{\includegraphics[angle=-90,width=9.0cm]{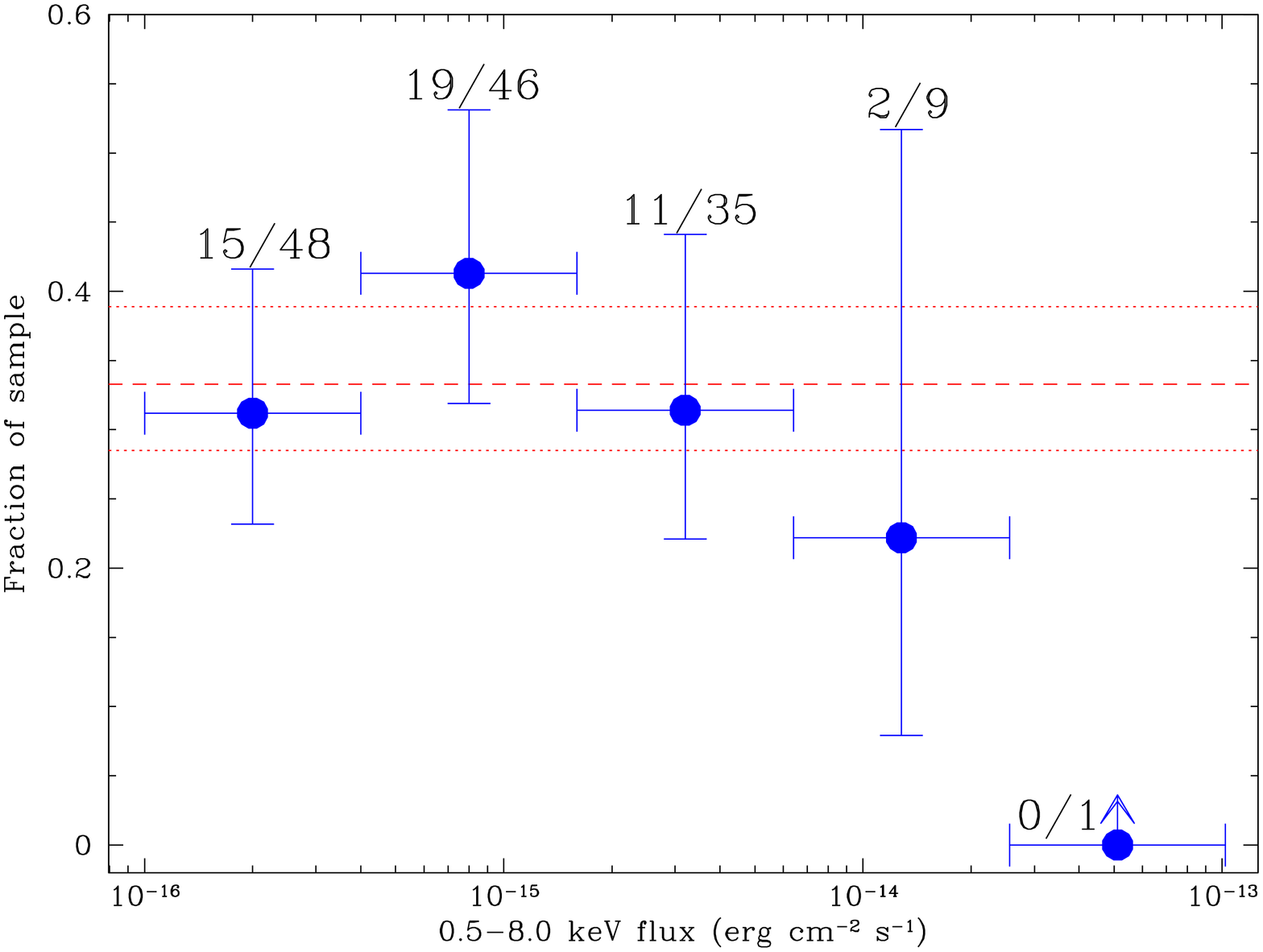}}
\figcaption{The fraction of optically faint X-ray sources versus full-band flux. The width of each X-ray flux bin is shown as bars in the $x$-axis direction. The 1$\sigma$ uncertainty in the fraction of optically faint X-ray sources is shown as bars in the $y$-axis direction. The numbers plotted for each data point show the number of optically faint X-ray sources over the total number of X-ray sources for each X-ray flux bin. The dashed line shows the overall fraction of optically faint X-ray sources in the entire X-ray sample, and the dotted lines show the 1$\sigma$ uncertainty on this value. The fraction of optically faint X-ray sources is consistent with being constant ($\approx 35$\%) for full-band fluxes $<3\times 10^{-14}$~erg~cm$^{-2}$~s$^{-1}$.}
\label{sourceexp}
\vspace{0.2in}

%%%%%%%%%%%%%%%%%%%%%%%%%%%%%%%%%%%%%%%%%%%%%%%%%%%%%%%%%%%%%%%%%%%%%%

\subsection{X-ray background contribution}

Due to the significant number of sources with upper limits in the soft or the hard X-ray bands, the most accurate determination of the X-ray background contribution from the optically faint X-ray sources is made by stacking the individual X-ray sources; see also Paper II and Paper IV for use of the stacking technique. The total number of counts for each source is measured in an aperture equal to the size of the point-spread function (see \S3.2.1 in Paper V); the size and shape of the point-spread function is a function of off-axis angle. The average number of background counts is calculated and removed to give the net number of counts per source. The results from all sources are combined to give a total number of counts for both the optically faint and optically bright X-ray samples. The total flux for each sample is then calculated using the average X-ray band ratio (see \S4.3), which was corrected for vignetting. To check this technique we determined the average band ratio of the whole sample; we found this to be $0.59\pm0.01$, which corresponds to $\Gamma\approx 1.32$, similar to that found for the spectral slope of the X-ray background (e.g.,\ compare to Tozzi \etal 2001). The optically faint X-ray sources contribute $\approx 22^{+8}_{-7}$\% of the total X-ray emission in the full band, $\approx 14^{+8}_{-7}$\% of the total X-ray emission in the soft band, and $\approx 24^{+8}_{-7}$\% of the total X-ray emission in the hard band. The uncertainties in these values have been determined assuming the limiting factor in these calculations is the small number of sources. We note, however, that ``cosmic variance'' is also likely to be important.

Based on the analysis of G.P.~Garmire \etal, in preparation, which uses the normalization of Chen, Fabian, \& Gendreau (1997), these HDF-N observations resolve $\approx 100$\% of the X-ray background in the soft band and $\approx 86$\% of the X-ray background in the hard band. Therefore, the optically faint X-ray source population contributes a non-negligible fraction ($\approx 14^{+8}_{-7}$\% in the soft band and $\approx 21^{+8}_{-7}$\% in the hard band) of the X-ray background.

%
%%%%%%%%%%%%%%%%%%%%%%%%%%%%%%%%%%%%%%%%%%%%%%%%%%%%%%%%%%%%%%%%%%%%%%
\section{Comparisons of the optically faint and optically bright X-ray source populations}
%%%%%%%%%%%%%%%%%%%%%%%%%%%%%%%%%%%%%%%%%%%%%%%%%%%%%%%%%%%%%%%%%%%%%%
%

In this section we provide constraints on the nature of the optically faint X-ray source population from a comparison to the optical, near-IR and X-ray properties of the optically bright X-ray source population.

\subsection{Optical-to-near-IR colors}

Deep X-ray surveys show a correlation between the optical faintness and optical-to-near-IR color of X-ray sources (e.g.,\ Hasinger \etal 1998; Giacconi \etal 2001; Lehmann \etal 2001). In Figure~6 we show a plot of $I-K$ color versus $I$-band magnitude for the entire X-ray sample; the $K$-band magnitude was determined from $K=HK^\prime-0.3$ following Barger \etal (1999). The correlation between the optical magnitude and optical-to-near-IR color of X-ray sources is clearly seen in our data. The Spearman $\rho$ and Kendall $\tau$ tests show a correlation is present with $>$99.99\% confidence for the optically bright sources; we omitted the optically faint sources due to the larger uncertainties in the source magnitudes. With the exception of the BLAGN, the X-ray sources are among the reddest sources at a given optical magnitude, and the best linear fit to the non-BLAGN sources is very similar to that found for the $\mu$Jy radio source population (Richards \etal 1999). Two obvious mechanisms can produce this reddening effect: (1) extinction of the optical continuum and/or, (2) the positive $K$-corrections of a normal galaxy with increasing redshift (e.g.,\ Pozzetti \& Mannucci 2000). The expected redshifts of elliptical and spiral galaxies with $I-K>3.5$ are $z>1.0$ and $z>1.5$ respectively (e.g.,\ Moriondo, Cimatti, \& Daddi 2000; Pozzetti \& Mannucci 2000; Barger \etal 2001a). The luminous NLAGN sources follow the trend found for the non-BLAGN sources showing that their optical-to-near-IR emission is probably dominated by the host galaxy. By comparison, the BLAGN have blue colors over a large range of redshifts ($0.5<z<3.5$ for our sample), presumably due to the domination of the AGN emission at these wavelengths.

The majority of the optically faint X-ray sources have red optical-to-near-IR colors ($I-K\ge3.5$), and nine (30$^{+14}_{-10}$\%) of the 30 sources with measurable colors are very red objects (VROs; $I-K\ge4$); by comparison only one (1$^{+2}_{-1}$\%) of the 94 optically bright X-ray sources is a VRO. A detailed analysis of the X-ray emission from VROs will be published elsewhere (D.M.~Alexander \etal, in preparation). In general these colors are inconsistent with those expected for a normal BLAGN at $z\simlt$~6 (i.e.,\ before Lyman-$\alpha$ leaves the $I$-band; see Figure~6 and \S6.2), although they are consistent with those expected for extremely high redshift BLAGN, reddened BLAGN (e.g.,\ Webster \etal 1995; Barkhouse \& Hall 2001), and comparitively normal galaxies that are either dust extincted and/or lie at $z>1$.

%
%%%%%%%%%%%%%%%%%%%%%%%%%%%%%%%%%%%%%%%%%%%%%%%%%%%%%%%%%%%%%%%%%%%%%%
% 6 Optical to near-IR colours for optical magnitude
%%%%%%%%%%%%%%%%%%%%%%%%%%%%%%%%%%%%%%%%%%%%%%%%%%%%%%%%%%%%%%%%%%%%%%
%

\vspace{0.2in}
\centerline{\includegraphics[angle=-90,width=9.0cm]{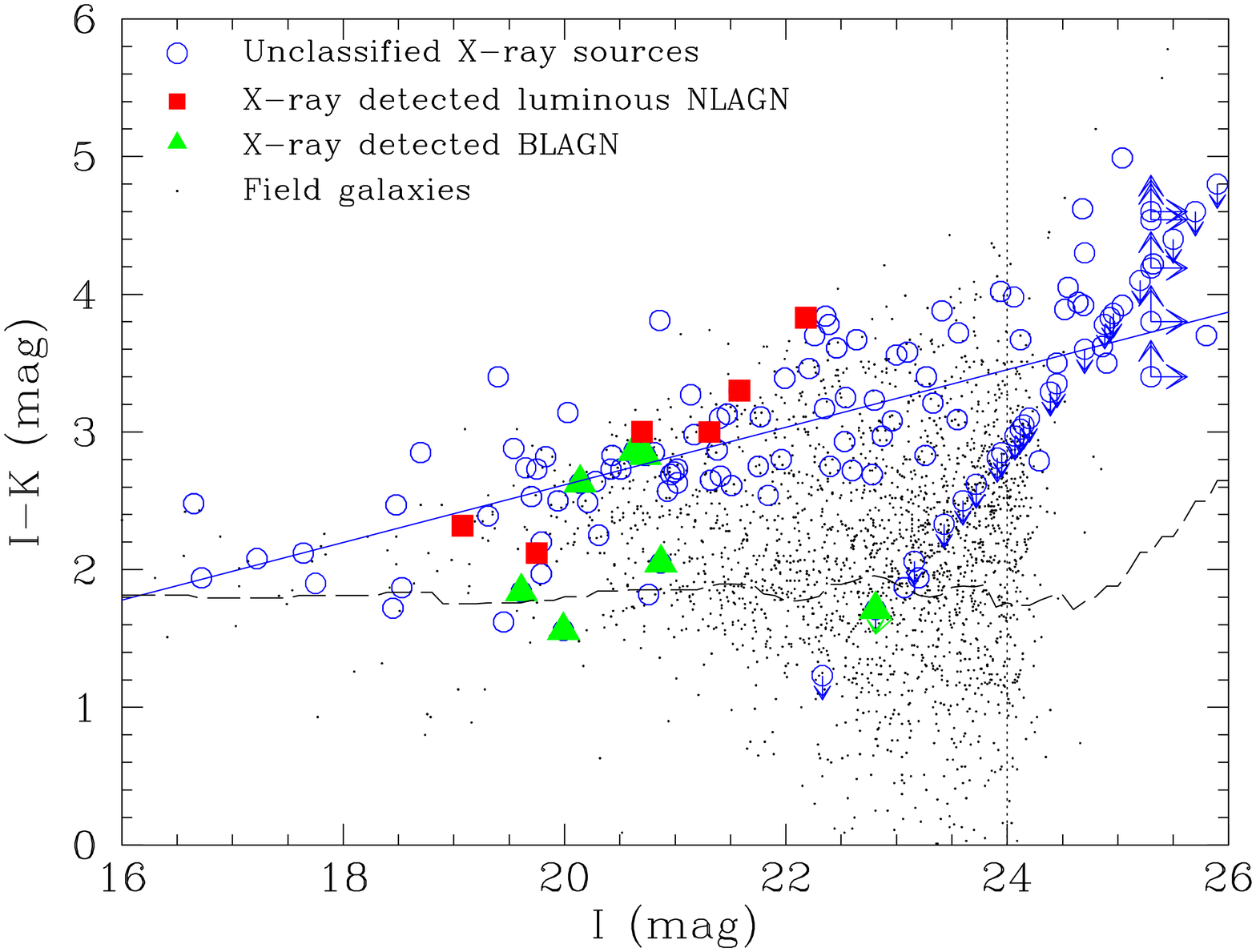}}
\figcaption{$I-K$ color versus $I$-band magnitude for the entire X-ray sample; the $K$-band magnitude was determined from $K=HK^\prime-0.3$ following Barger \etal (1999). The small dots are the field galaxy population (from Barger \etal 1999), the open circles are the unclassified \chandra\ sources, the filled triangles are the BLAGN sources, and the filled squares are the luminous NLAGN. The solid line shows the best linear fit to the non-BLAGN sources; this is comparable to the slope for the $\mu$Jy radio sources shown in Figure~2 of Richards \etal (1999). The Spearman $\rho$ and Kendall $\tau$ tests show a correlation is present with $>$99.99\% confidence. The BLAGN do not follow this trend; their average $I-K$ color stays roughly constant for $I<24$. The long dashed line shows the expected colors for a $M_{I}=-23$ QSO. The QSO colors were calculated from a standard quasar spectrum with $\alpha=0.5$ (where $F_{\nu}\propto\nu^{-\alpha}$) and typical emission-line strengths. The colors correspond to that expected for an $M_{I}=-23$ QSO, see Figure~12 for the equivalent redshifts; at $z\simgt 6$, Lyman-$\alpha$ leaves the $I$-band leading to large $I-K$ colors. The vertical dotted line shows the optical magnitude distinction between optically bright ($I<24$) and optically faint ($I\ge24$) sources. The diagonal line of $I-K$ color upper limits is due to the $HK^\prime$ magnitude limit.}
\label{fig:imk}
\vspace{0.2in}

%%%%%%%%%%%%%%%%%%%%%%%%%%%%%%%%%%%%%%%%%%%%%%%%%%%%%%%%%%%%%%%%%%%%%%

\subsection{X-ray-to-optical flux ratios}

An important diagnostic of the nature of X-ray sources is the X-ray-to-optical flux ratio (e.g.,\ Maccacaro \etal 1988; Stocke \etal 1991). Luminous AGN (both BLAGN and NLAGN) have typical X-ray-to-optical flux ratios, in both the soft and hard bands, of $-1<\log{({{f_{\rm X}}\over{f_{\rm R}}})}<1$ (e.g.,\ Schmidt \etal 1998; Akiyama \etal 2000; Paper II; Lehmann \etal 2001). A large fraction of sources are also detected in the soft band with lower X-ray-to-optical flux ratios (i.e.,\ $\log{({{f_{\rm X}}\over{f_{\rm R}}})}<-1$). These sources include normal galaxies, stars and low-luminosity AGN (e.g.,\ Giacconi \etal 2001; Paper II; Lehmann \etal 2001; Paper IV; A.E. Hornschemeier \etal, in preparation); normal galaxies and stars generally have weak hard X-ray emission and consequently very small X-ray-to-optical ratios (i.e.,\ $\log{({{f_{\rm X}}\over{f_{\rm R}}})}<-2$).

Previous studies have used either the $V$-band or $R$-band when determining the X-ray-to-optical flux ratios of sources whereas our study uses the $I$-band magnitude (see Figure~7). The average $R-I$ of the $I<24$ X-ray sources reported in Paper II is 0.9 mags, which corresponds to a difference between ${{f_{\rm X}}\over{f_{\rm I}}}$ and ${{f_{\rm X}}\over{f_{\rm R}}}$ of only $\approx 20$\% once the zero points of the $I$-band and $R$-band magnitude scales are accounted for. The optically bright BLAGN and luminous NLAGN sources lie within the typical range of X-ray-to-optical flux ratios found for luminous AGN (see Figure~7). The other optically bright sources cover a large range of soft-band X-ray-to-optical flux ratios, and, although we have not classified all these sources with optical spectroscopy, a
%
%%%%%%%%%%%%%%%%%%%%%%%%%%%%%%%%%%%%%%%%%%%%%%%%%%%%%%%%%%%%%%%%%%%%%%
% 7 X-ray-to-optical flux ratios
%%%%%%%%%%%%%%%%%%%%%%%%%%%%%%%%%%%%%%%%%%%%%%%%%%%%%%%%%%%%%%%%%%%%%%
%
\vspace{0.2in}
\centerline{\includegraphics[angle=0,width=9.0cm]{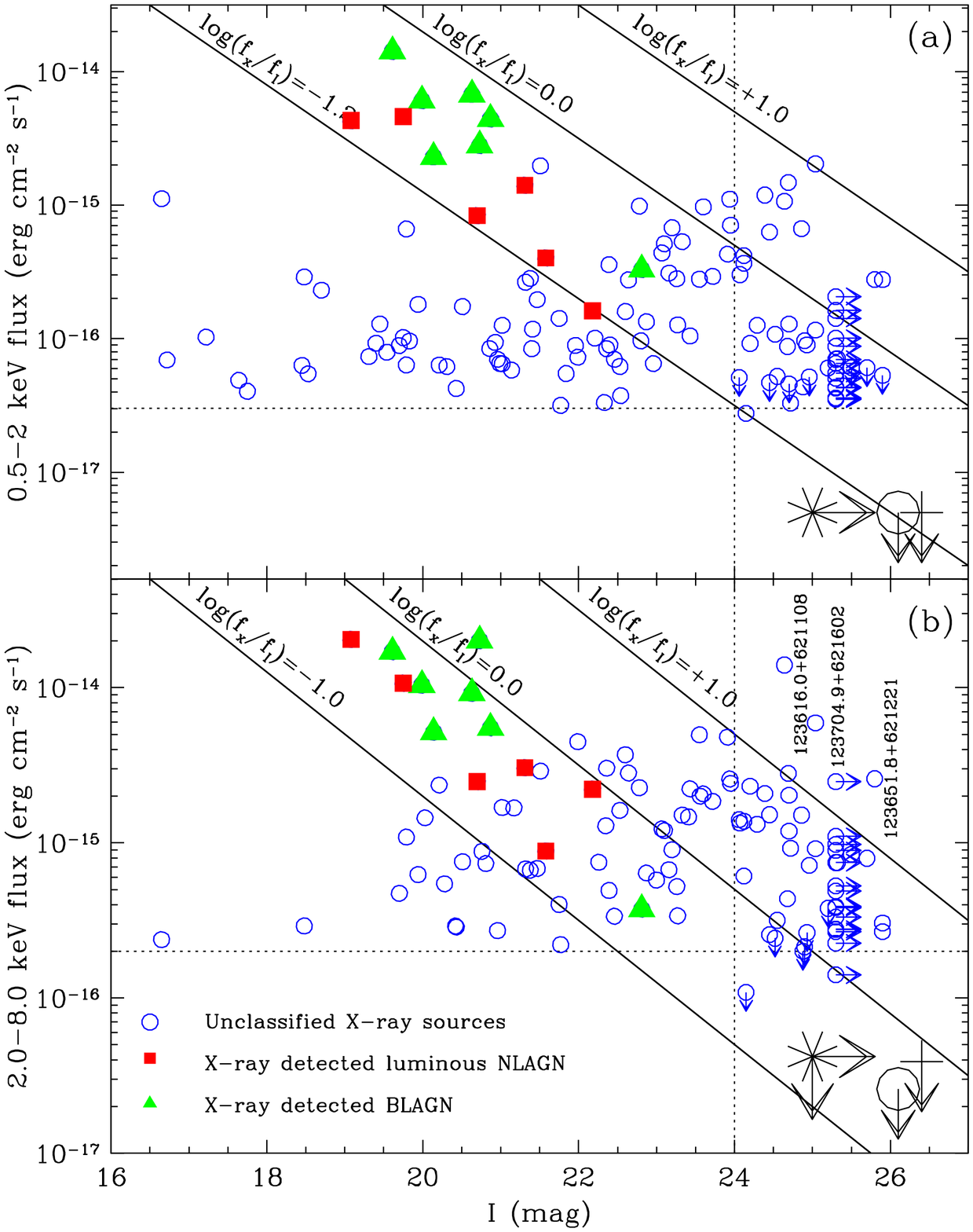}}
\vspace{0.0in}
\figcaption{(a) Soft-band and (b) hard-band fluxes versus $I$-band magnitude; both individual \chandra\ sources and the results of stacking sources not individually detected by \chandra\ (see \S5.1 and \S5.2) are shown. The optically bright X-ray sources with upper limits are not plotted; both X-ray detections and X-ray upper limits are plotted for the optically faint X-ray sources. The dotted lines in the horizontal and vertical directions show the X-ray flux limits and the distinction between optically bright and optically faint sources respectively. The small open circles are the unclassified sources, the filled triangles are the BLAGN, and the filled squares are luminous NLAGN. The large cross is the stacking analysis limit for the Conti \etal (1999) optical AGN candidates, the large open circle is the stacking analysis limit for the Jarvis \& MacAlpine (1998) optical AGN candidates, and the large star is the stacking analysis limit for the Richards \etal (1999) optically faint $\mu$Jy radio sources. Three sources with extreme flux ratios are labeled. The source CXOHDFN J123651.8+621221 is likely to be an obscured QSO (see \S6.2).}
\label{fig:oxratios}
\vspace{0.5in}

%%%%%%%%%%%%%%%%%%%%%%%%%%%%%%%%%%%%%%%%%%%%%%%%%%%%%%%%%%%%%%%%%%%%%%

\noindent broad range of source types are detected (see Figure~7a).

It is clear that the optical magnitude threshold of the optically faint X-ray sample restricts the range of possible source types for this population, and we have probably not yet reached the X-ray sensitivity needed to detect optically faint non-AGN sources (see \S5.1 for some possible exceptions). This point is further enforced in Figure~7b which shows that the optically faint sources have hard-band X-ray-to-optical flux ratios typical of AGN. Some optically faint sources are among the brightest X-ray sources in the entire X-ray sample, suggesting luminous AGN activity even at moderate redshift. For example, the two brightest X-ray sources (CXOHDFN J123616.0+621108 and CXOHDFN J123704.9+621602) would have QSO-level X-ray luminosities (rest-frame 0.5--8.0~keV unabsorbed luminosities of $>3\times10^{44}$ ergs s$^{-1}$) at $z\approx 1.5$; see \S6.3.

\subsection{X-ray band ratios}

One of the key distinctions between the main classes of AGN is made from X-ray observations. Unobscured AGN are almost always BLAGN and have steep X-ray spectral slopes (e.g.,\ $\Gamma=2.0\pm0.3$; George \etal 2000), while obscured AGN are predominantly NLAGN and have highly absorbed X-ray emission (e.g.,\ $22<\log(N_{\rm H})<25$; see Risaliti, Maiolino, \& Salvati 1999) and consequently flat X-ray spectral slopes (e.g.,\ $\Gamma\approx 1.0$). While AGN with steep X-ray spectral slopes are almost exclusively BLAGN, a number of source types other than NLAGN can have flat X-ray spectral slopes [e.g.,\ Broad Absorption Line QSOs (BALQSOs) and other obscured Type 1 AGN; Gallagher \etal 1999; Comastri \etal 2001]. A comparison of the X-ray spectral slopes of the optically faint X-ray sources and the optically bright X-ray sources will therefore provide constraints on whether the majority of the AGN activity in the optically faint X-ray source population is obscured or unobscured.

In Figure~8 we show a plot of the X-ray band ratio, defined as the ratio of hard-band to soft-band counts, versus the soft-band count rate. The general trend toward flatter X-ray spectral slopes at fainter X-ray fluxes (e.g.,\ Giacconi \etal 2001; Papers II--V; Tozzi \etal 2001) is seen. Clearly the optically faint and optically bright X-ray source populations have a distribution of soft and hard X-ray sources. To determine whether the optically faint sources have statistically flatter X-ray spectral slopes we used two techniques. First we stacked together the individual X-ray source observations in the same manner as in \S3.3. The average band ratios from this stacking analysis, corrected for vignetting, are $0.91\pm0.03$ ($\Gamma=0.9$) for the optically faint sample and $0.54\pm0.01$ ($\Gamma=1.4$) for the optically bright sample. The flat X-ray spectral slope of the optically faint X-ray sources suggests the majority of these sources are obscured AGN; by comparison the average band ratio for the luminous NLAGN is $0.68\pm0.02$ ($\Gamma=1.2$). Second we compared the band ratio distribution of the optically faint X-ray sources to the band ratio distribution of the optically bright X-ray sources (see Figure~9). It is clear from this plot that there is a larger fraction of optically faint sources with flat X-ray spectral slopes than found in the optically bright X-ray source population. The K-S test shows that these band ratio distributions are distinguishable at the 99.4\% significance level. Assuming the underlying emission is an unabsorbed power-law source with $\Gamma=2.0$, the average band ratio of the optically faint X-ray sources corresponds to an intrinsic absorption column density at $z=2$ (see \S6.2) of \hbox{$N_{\rm H}\approx 1.5\times10^{23}$ cm$^{-2}$}. Although it is probable that the majority of these sources are NLAGN, without optical spectroscopic observations we cannot distinguish between NLAGN and other X-ray sources that can have flat X-ray spectral slopes (e.g.,\ BALQSOs and other obscured Type 1 AGN).

Without high signal-to-noise X-ray spectral analysis we cannot directly show that the flattening of the X-ray spectral slopes in the optically faint X-ray source population is due to absorption. While in principle the signal-to-noise ratio of the X-ray emission from the stacked optically faint sources is high enough to allow spectral analysis, the probable broad range of source redshifts (see \S6.2) will smear out the signature of absorption and other X-ray spectral features (e.g.,\ iron K$\alpha$ lines). We also note that when determining the average band ratios we have not made $K$-corrections to the X-ray emission. As the effect of 
%
%%%%%%%%%%%%%%%%%%%%%%%%%%%%%%%%%%%%%%%%%%%%%%%%%%%%%%%%%%%%%%%%%%%%%%
% 8 Hardness ratios
%%%%%%%%%%%%%%%%%%%%%%%%%%%%%%%%%%%%%%%%%%%%%%%%%%%%%%%%%%%%%%%%%%%%%%
%
\vspace{0.2in}
\centerline{\includegraphics[angle=-90,width=9.0cm]{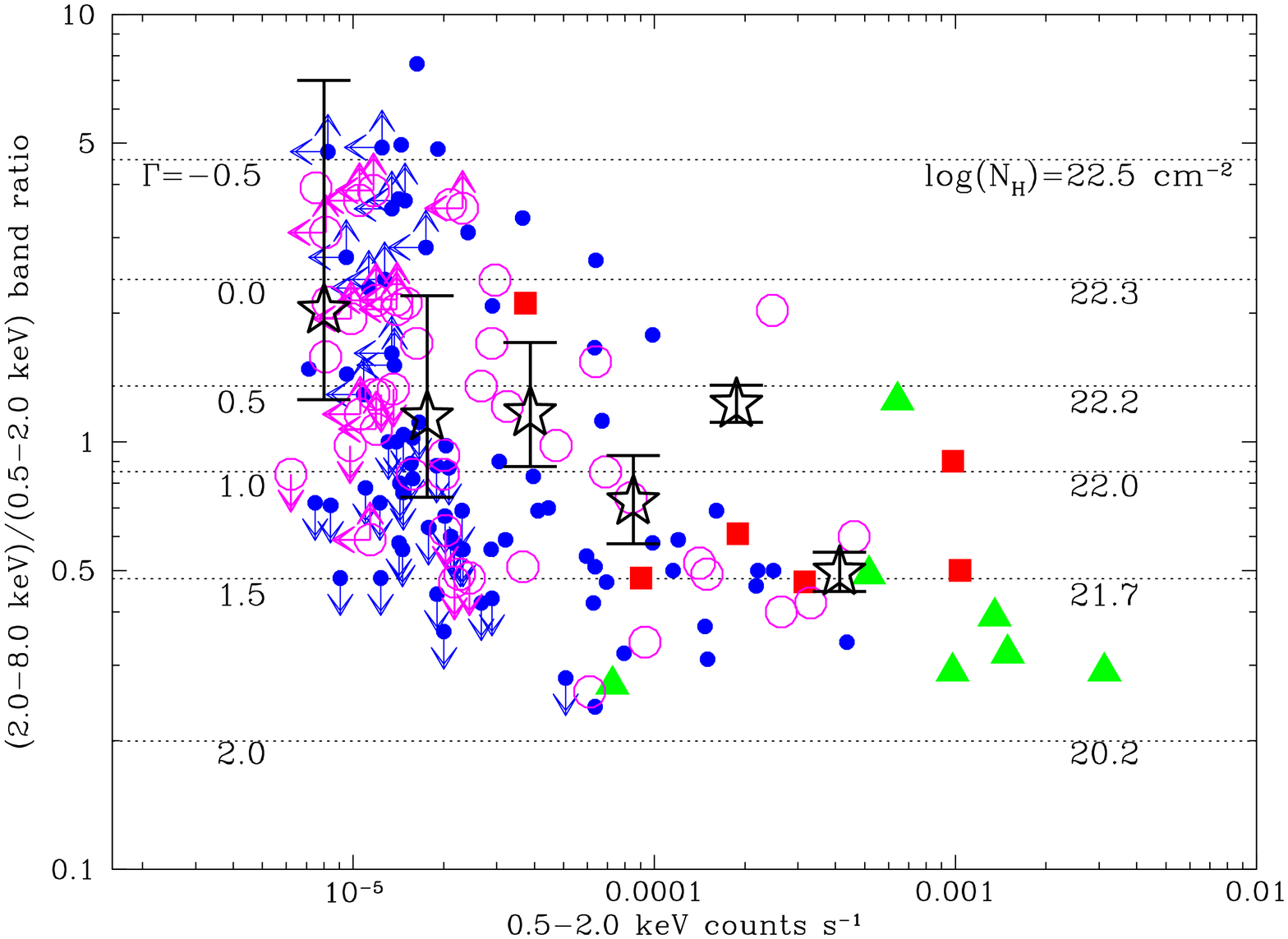}}
\figcaption{X-ray band ratio, defined as the ratio of hard-band to soft-band counts, versus soft-band count rate. The open circles are the optically faint sources, the filled circles are the optically bright sources, the filled triangles are the BLAGN sources, and the filled squares are the luminous NLAGN. The large stars show the average band ratios for different soft-band count rates derived from stacking analyses of the optically faint X-ray sources (compare to Figure~12 in Paper V). The error bars on each stacked band ratio show the average size of the errors for a source of the given soft-band count rate; to reduce symbol crowding individual error bars have not been plotted. The equivalent photon indices ($\Gamma$) are shown on the left-hand side of the figure, and the equivalent column densities ($N_H$) for a $\Gamma$=~2.0 power-law source at $z\approx 0$ are shown on the right-hand side of the figure. These values were determined using the AO2 version of {\sc pimms} (Mukai 2000). A trend toward flatter X-ray spectral slopes for fainter soft-band fluxes is seen (compare to Paper V).}
\label{fig:hardness}
\vspace{0.2in}

%%%%%%%%%%%%%%%%%%%%%%%%%%%%%%%%%%%%%%%%%%%%%%%%%%%%%%%%%%%%%%%%%%%%%%

\noindent redshifting sources of similar absorbing column densities leads to steeper X-ray spectral slopes at higher redshift, the probable higher redshifts of the optically faint X-ray sources (see \S6.2) suggests that their intrinsic absorbing column densities are considerably higher than those found for the luminous NLAGN. A more detailed comparison is difficult as \chandra\ is more sensitive to absorbed sources at high redshift than at low redshift, and therefore poorly understood selection effects would also need to be considered.

\subsection{X-ray variability}

Another key signature of AGN activity is X-ray variability on timescales of minutes to years (e.g.,\ Mushotzky, Done, \& Pounds 1993). As this study concerns several observations taken over a 16-month period (see Paper V for observation dates and exposure times), we can test whether these sources show evidence for X-ray variability. A useful tool for detecting X-ray variability is the K-S test (see Paper IV). Variability was assessed in all the X-ray bands. We guarded against false variability by fitting a constant model to the data points of each source and accepting only those sources where the $\chi^2$ fit is rejected with $\ge$90\% confidence. The source event extraction radius was taken to be twice the point spread function size to ensure that all the source counts were included in the testing. To further reduce the possibility of spurious detections of variability, we only included those sources with more than 100 full-band counts and where the K-S test showed evidence for X-ray variability at the $>$99.5\% confidence level in at least one X-ray band. We would expect 0.14 spurious detections of variability over all bands at this confidence level. Four of the nine ($44^{+35}_{-21}$\%) optically faint 
%
%%%%%%%%%%%%%%%%%%%%%%%%%%%%%%%%%%%%%%%%%%%%%%%%%%%%%%%%%%%%%%%%%%%%%%
% 9 Hardness ratio distributions
%%%%%%%%%%%%%%%%%%%%%%%%%%%%%%%%%%%%%%%%%%%%%%%%%%%%%%%%%%%%%%%%%%%%%%
%

\vspace{0.2in}
\centerline{\includegraphics[angle=0,width=9.0cm]{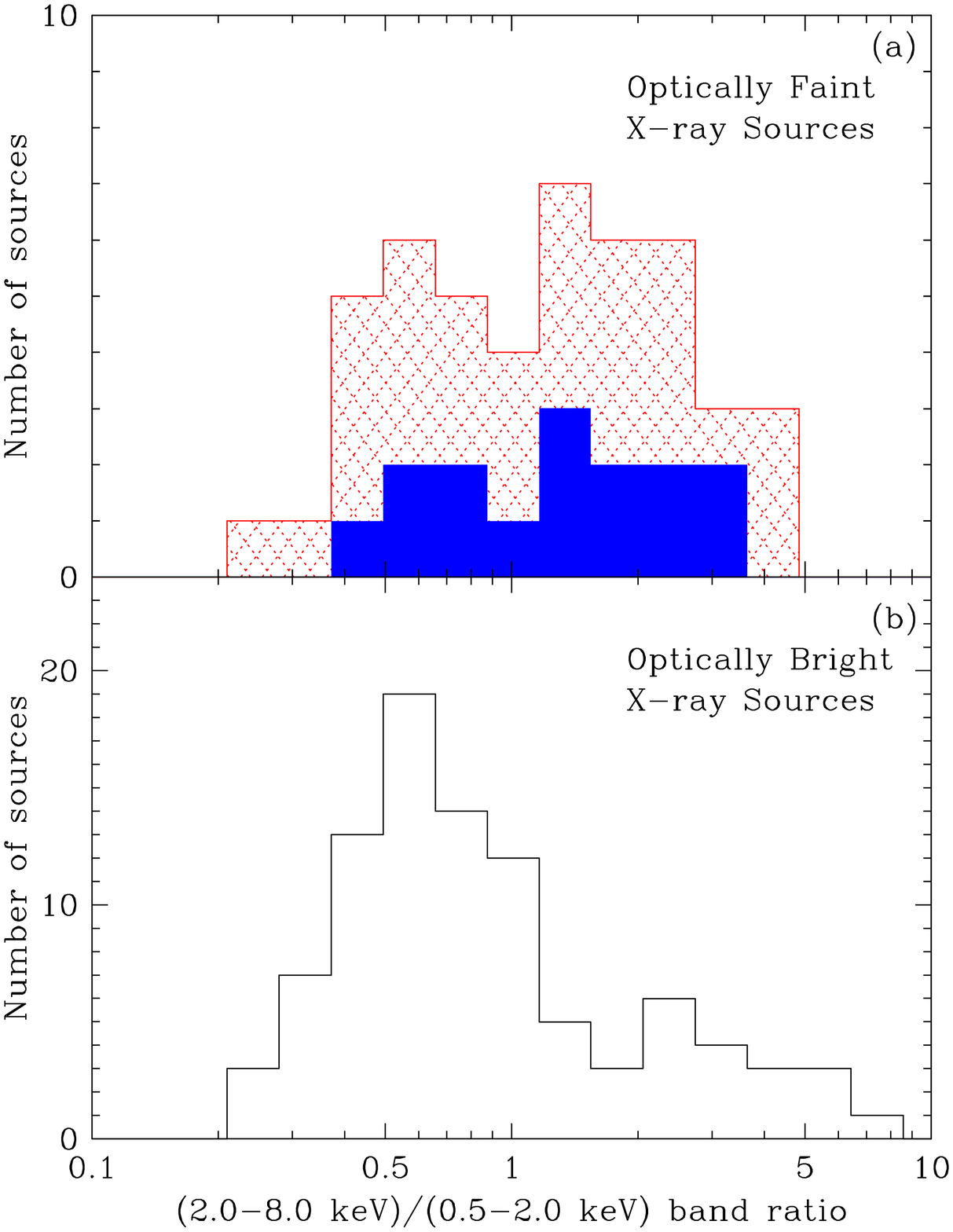}}
\figcaption{X-ray band ratio distributions of (a) the optically faint sample and (b) the optically bright sample. Clearly a larger fraction of optically faint X-ray sources have flat X-ray spectral slopes. The two X-ray band ratio distributions are distinguishable according to the K-S test at the 99.4\% significance level. The solid blocks show the overlaid band ratio distribution of sources without 2$\sigma$ $I$-band counterparts.}
\label{sourceexp}
\vspace{0.2in}

%%%%%%%%%%%%%%%%%%%%%%%%%%%%%%%%%%%%%%%%%%%%%%%%%%%%%%%%%%%%%%%%%%%%%%

\noindent X-ray sources matching this criteria showed evidence for variability (see Figure~10). The evidence for variability in all of these sources is at the $>$99.9\% level in the soft and full bands and at the $>$99\% level in the hard band, with the exception of CXOHDFN J123722.7+620935 which shows evidence for hard band variability at the 95.4\% confidence level. Of the three sources in Figure~7b with large hard band X-ray-to-optical flux ratios (i.e.,\ $\log{({{f_{\rm X}}\over{f_{\rm I}}})}>>1$), only CXOHDFN J123651.8+621221 does not show evidence for variability. All of the X-ray variable sources have $I$-band counterparts and, with the exception of CXOHDFN J123722.7+620935, red optical-to-near-IR colors (i.e.,\ $I-K>3.5$).

%
%%%%%%%%%%%%%%%%%%%%%%%%%%%%%%%%%%%%%%%%%%%%%%%%%%%%%%%%%%%%%%%%%%%%%%
% 10 Lightcurves I
%%%%%%%%%%%%%%%%%%%%%%%%%%%%%%%%%%%%%%%%%%%%%%%%%%%%%%%%%%%%%%%%%%%%%%
%

\begin{figure*}
\vspace{0.0in}
\centerline{\includegraphics[angle=0,width=18.5cm]{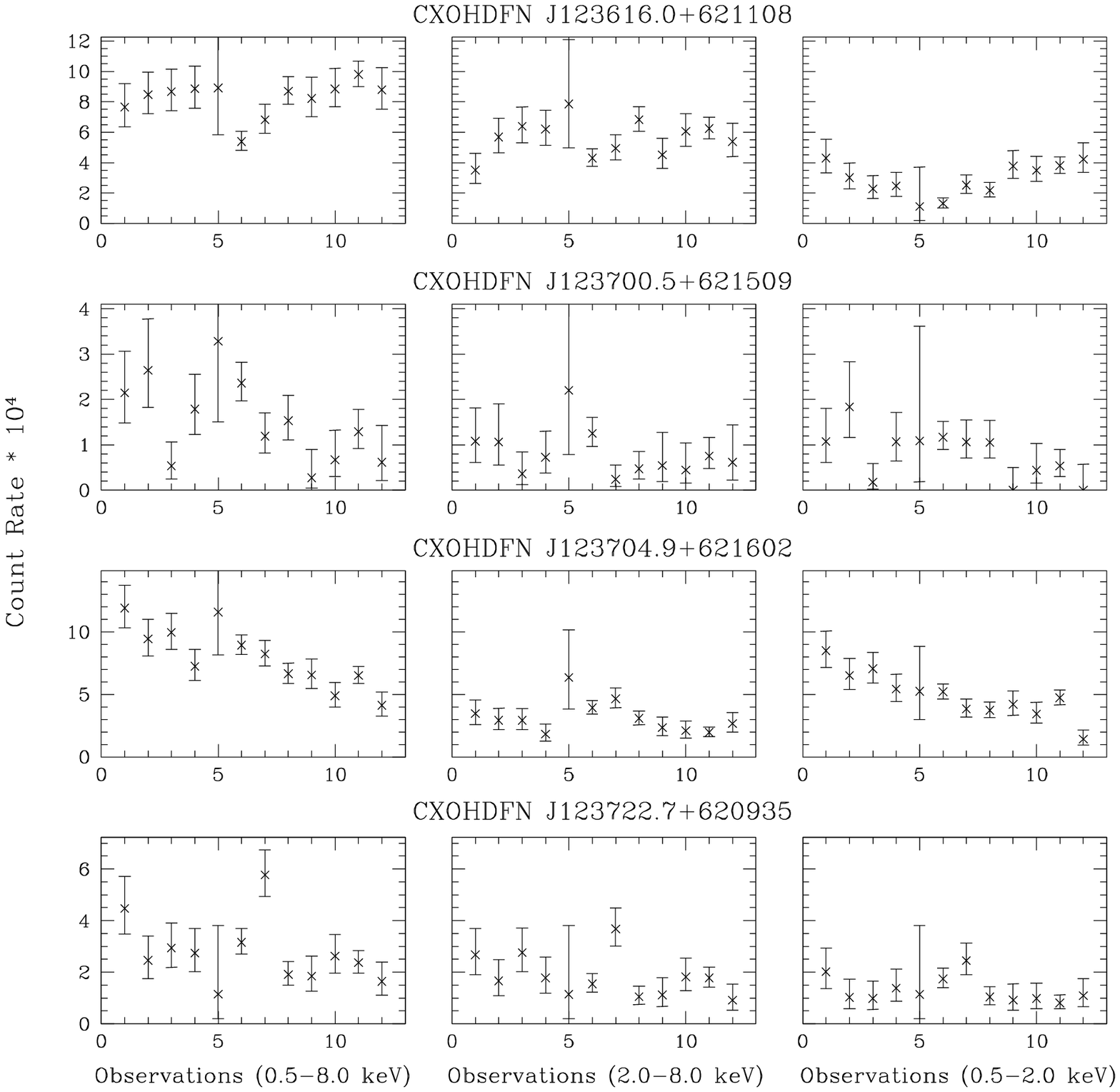}}
\vspace{0.5in}
\figcaption{\chandra\ light curves for optically faint X-ray sources showing evidence of X-ray variability. The average count rate is shown for each of the individual \chandra\ observations. The observations are listed in Table 1 of Paper V. The observations vary in length, and {\bf the observation spacings are not constant}. Error bars have been calculated following Gehrels (1986). The confidence level for variability in all of the sources is $>$99.5\% in at least one X-ray band; see \S4.4 for further discussion.}
%\epsscale{0.7}
\label{fig:lightcurves}
\end{figure*}

%%%%%%%%%%%%%%%%%%%%%%%%%%%%%%%%%%%%%%%%%%%%%%%%%%%%%%%%%%%%%%%%%%%%%%

The K-S test is more sensitive to detecting variability in bright sources. However, since the full-band flux distributions of the optically faint and optically bright X-ray samples are statistically consistent, we can compare the fraction of variable sources in both samples using the above criteria without undue bias. Within the optically bright X-ray sample, 13 ($45^{+16}_{-12}$\%) of the 29 sources with more than 100 full band counts showed evidence for variability with the above criteria; we would expect 0.4 spurious detections of variability over all bands at this confidence level. All (i.e.,\ 100\%) of the BLAGN and two ($33^{+44}_{-21}$\%) of the six luminous NLAGN with more than 100 full band counts were found to be variable. If the BLAGN and luminous NLAGN sources are removed from the optically bright X-ray sample then 5 ($29^{+20}_{-13}$\%) of the 17 sources showed evidence for variability. Therefore, although the statistics are limited, the lower fraction of variable sources in the optically faint X-ray sample (i.e.,\ $44^{+35}_{-21}$\%) is more consistent with the majority of the optically faint X-ray source population being luminous NLAGN rather than BLAGN. These results are in good agreement with the optical-to-near-IR color (\S4.1) and X-ray band ratio analyses (\S4.3).

\subsection{X-ray emission from galaxy clusters}

We have not yet considered the possibility that the X-ray emission from some optically faint X-ray sources is cluster emission. Clusters can have X-ray-to-optical flux ratios as large as those observed for some of our optically faint X-ray sources (e.g., Stocke \etal 1991). Furthermore, the X-ray centroids of some clusters are offset from their optical centroids which can lead to even larger {\it apparent} X-ray-to-optical flux ratios. We searched the optical images around our optically faint X-ray sources, and none of the optically faint X-ray sources reside in an obvious cluster. Furthermore, none of the optically faint X-ray sources show evidence for extended X-ray emission.\footnote{The optically faint X-ray point sources CXOHDFN J123621.9+621603, CXOHDFN J123705.1+621635, and CXOHDFN J123618.4+621551 (also a $\mu$Jy radio source; see \S5.1) do however lie within diffuse X-ray emission (see F.E.~Bauer \etal, in preparation).} Finally, the generally flat X-ray spectra of the optically faint X-ray sources are inconsistent with the soft thermal X-ray emission produced by clusters. According to the luminosity-temperature relationship for clusters (e.g., Wu, Xue, \& Fang 1999), a low-redshift cluster would need to have an X-ray luminosity of $>3\times 10^{44}$~erg~s$^{-1}$ in order to produce emission as hard as that observed from our softest optically faint X-ray sources. However, given this luminosity, a low-redshift cluster would have an X-ray flux much larger than observed from any of our optically faint X-ray sources. A cluster at higher redshift would need an even higher temperature (due to the redshifting of the X-ray spectrum) and thus an even greater luminosity according to the luminosity-temperature relationship; there is no self-consistent solution where a cluster can give both the observed band ratio and the observed X-ray flux for an optically faint X-ray source.

\subsection{X-ray emission from optically faint Galactic objects}

We have only considered the possibility that our optically faint X-ray sources are extragalactic objects, as it is unlikely that the optically faint X-ray source population is significantly contaminated by Galactic objects. Considering first normal stars, only extreme M stars could have X-ray-to-optical flux ratios as large as those observed (e.g., Maccacaro et~al. 1988), however the generally large band ratios of our optically faint X-ray sources are not consistent with those expected for M stars. Isolated neutron stars can have extremely large X-ray-to-optical flux ratios (e.g.,\ Treves \etal 2000) and would appear as optically blank sources at the depth of our X-ray survey; however, we would expect only $\approx$~0.1 detectable isolated neutron stars in our field (e.g.,\ Popov \etal 2000). Due to the fact that our survey is much deeper than those performed previously, it is difficult to rule out rigorously some contamination by a new, previously unknown class of Galactic object. However, known Galactic objects with large X-ray-to-optical flux ratios and hard X-ray spectra, such as low-mass X-ray binaries and cataclysmic variables, have much lower densities on the sky than observed for our optically faint X-ray sources (e.g.,\ Howell \& Szkody 1990; van~Paradijs 1995), especially given the high Galactic latitude of $b=54\fdg 8$ for our field.

%
%%%%%%%%%%%%%%%%%%%%%%%%%%%%%%%%%%%%%%%%%%%%%%%%%%%%%%%%%%%%%%%%%%%%%%
\section{X-ray emission from other optically faint source populations}
%%%%%%%%%%%%%%%%%%%%%%%%%%%%%%%%%%%%%%%%%%%%%%%%%%%%%%%%%%%%%%%%%%%%%%
%

In \S4, we considered the nature of optically faint X-ray sources. In this section, we investigate the X-ray properties of two other optically faint source populations: optically faint $\mu$Jy radio sources and optically faint AGN candidate sources.

\subsection{X-ray emission from the optically faint \boldmath$\mu$Jy radio source population}

There are 17 optically faint $\mu$Jy radio sources lying within the region of our study (Richards \etal 1999). We found six of these sources to be positionally coincident to within 1\asec\ of optically faint X-ray sources. We ran {\sc wavdetect} (Freeman \etal 2001) with a probability threshold of 10$^{-5}$ over the positions of the undetected optically faint $\mu$Jy radio sources and detected a further 3 sources, see Table~2; we would expect 0.001 spurious sources in each X-ray band at this detection threshold. Three of these nine X-ray sources have been previously reported (see Table~2; Paper I; Paper II; Paper IV), and their flat X-ray spectral slopes and/or luminous X-ray emission suggest AGN activity. Of the 6 previously unreported sources, two have flat radio spectra (VLA J123707+621408; VLA J123721+621130) suggesting AGN activity (see Table~2; Richards 1999). The radio properties of the other sources are consistent with either AGN or starburst activity. With the exception of VLA J123651+621221 and VLA J123707+621408, the X-ray detected optically faint $\mu$Jy radio sources have less than 25 full band counts, limiting the scope of X-ray analysis. However, we can stack the individual detections to provide a statistical measure of the X-ray spectral slope using the stacking technique described in \S3.3. Excluding the two brighter sources mentioned above (both of which have good evidence for AGN activity), the stacked band ratio of the other 7 X-ray detected sources is found to be $0.97^{+0.18}_{-0.16}$, corresponding to $\Gamma =0.8$. This band ratio is similar to that found for the optically faint X-ray source population (see \S4.3) and suggests that the origin of the X-ray emission in the majority of the X-ray detected optically faint $\mu$Jy radio sources is obscured AGN activity. This further suggests that the X-ray detected optically faint $\mu$Jy radio sources are the radio-bright analogs of the optically faint X-ray source population, and slightly deeper radio observations should uncover a significanly larger fraction of the optically faint X-ray source population.

Due to the extreme faintness of optically faint $\mu$Jy radio sources, little is known of their nature. Richards \etal (1999) proposed that the source population could be composed of three main source types: (1) luminous dust-enshrouded starburst systems at $z\approx$~1--3, (2) luminous obscured AGN at $z\simgt$~2, or (3) extreme redshift ($z>6$) AGN. Sub-millimeter observations have indeed shown that a large fraction ($\approx 50$\%) of this radio population appear to host dusty starbursts at $z\approx$~1--3 (e.g.,\ Barger, Cowie, \& Richards 2000; Chapman \etal 2001). Our observations have shown that a large fraction ($\approx 53^{+24}_{-17}$\%) of the population also has detectable X-ray emission. Whilst we have shown that the majority of the X-ray detected optically faint $\mu$Jy radio sources host obscured AGN activity, the detection of sub-millimeter emission in two of the X-ray detected sources (VLA J123618+621550 and VLA J123646+621448; see Barger, Cowie, \& Richards 2000) raises the question as to whether the X-ray emission in these two sources is due to luminous star formation activity. The soft-band fluxes of these sources are consistent with those expected from Arp~220 (e.g.,\ Iwasawa \etal 2001), the archetypal dusty starburst galaxy, at $z\approx 0.7$ or from NGC~3256 (Moran, Lehnert, \& Helfand 1999; Lira \etal 2001), the most X-ray luminous local starburst galaxy, at $z\approx 1.5$. At the millimetric redshifts of these sources ($z=$~$1.8^{+0.7}_{-0.6}$; VLA J123618+621550 and $z=$~$2.3^{+0.8}_{-0.7}$; VLA J123646+621448; see Barger \etal 2000), any star formation emission at X-ray energies would have to be at least as luminous as that found in NGC~3256.

There are 8 optically faint $\mu$Jy radio sources in the Richards \etal (1999) catalog with no detectable X-ray emission. We can search for evidence of X-ray emission by stacking the individual sources in the same manner as was performed for the faint X-ray detected sources above. The results of the stacking analysis are given in \hbox{Table~3}. A possible detection (at the 98.3\% confidence level) is found in the soft-band, corresponding to an average source flux of $5\times10^{-18}$~erg~cm$^{-2}$~s$^{-1}$, assuming $\Gamma=2.0$. Significant detections are not found in the full and hard bands, giving 3$\sigma$ upper limits of $2.2\times10^{-17}$~erg~cm$^{-2}$~s$^{-1}$ and $4.2\times10^{-17}$~erg~cm$^{-2}$~s$^{-1}$, respectively. The detected soft-band emission may be produced by AGN activity corresponding to rest-frame 0.5--2.0~keV luminosities of \hbox{$3\times10^{41}$ erg s$^{-1}$} at $z=3$ and \hbox{$2\times10^{42}$ erg s$^{-1}$} at $z=6$. However, a large fraction of this emission may be produced by star formation activity. This average soft band flux corresponds to that expected from Arp~220 at $z\approx 1.5$ or NGC~3256 at $z\approx 3.0$; compare to the results found by Brandt \etal (2001c, hereafter Paper VII) for Lyman-break galaxies. These redshifts bracket those found using the millimetric technique ($z=2$; e.g.,\ Chapman \etal 2001) although without deeper optical, near-IR and X-ray observations we cannot distinguish between AGN and star formation scenarios.

\subsection{X-ray constraints on the optically selected AGN candidates in the HDF-N}

At the X-ray flux limit of our survey, the surface density of the optically faint X-ray source population is $\approx 2,400^{+400}_{-350}$ deg$^{-2}$. Given the number density of $24<I<26$ sources in the \hbox{HDF-N} (e.g.,\ Fern\'andez-Soto, Lanzetta, \& Yahil 1999), we detect X-ray emission from $\approx 1.0$\% of the optically faint source population; this should be considered an upper limit as a fraction of our sources probably have $I>26$. The estimated fraction of optically selected AGN within the field galaxy population ranges from at least $\approx$~2--10\% (e.g.,\ Huchra \& Burg 1992; Tresse \etal 1996; Ho \etal 1997; Hammer \etal 1997), suggesting that a large fraction of optically faint AGN have not yet been detected in our \chandra\ observation. Within the HDF-N itself, Jarvis \& MacAlpine (1998) identified 12 candidate optically faint high-redshift AGN (see their Table~1) and Conti \etal (1999) identified 8 candidate optically faint moderate-redshift AGN (see their Table~4). None of these sources have been detected individually with X-ray emission (see also Paper II); however, as these sources are extremely faint [$F814W=26.1\pm0.7$ for the Jarvis \& MacAlpine (1998) sources and $F814W=26.4\pm0.3$ for the Conti \etal (1999) sources], the upper limits on their X-ray-to-optical flux ratios are still consistent with AGN activity (see Figure~7). 

We can place tighter constraints on their X-ray emission properties by stacking the individual sources in the same manner as was performed for the optically faint $\mu$Jy radio sources in \S5.1. The results of the stacking analyses are given in Table 3.\footnote{The Conti \etal (1999) source 94 lies too close to an X-ray source to be used in the stacking analysis.} We do not obtain a significant detection in any X-ray band for either of the candidate source lists. Assuming a typical AGN X-ray power-law of $\Gamma=2.0$, the 3$\sigma$ soft-band upper limits are $\approx 5\times10^{-18}$ ~erg~cm$^{-2}$~s$^{-1}$ for both the Jarvis \& MacAlpine (1998) and Conti \etal (1999) candidates. We can place constraints on the average luminosity of these sources as they all have photometric redshifts in the Fern\'andez-Soto, Lanzetta, \& Yahil (1999) catalog. The average redshifts are $z=3.4\pm1.4$ for the Jarvis \& MacAlpine (1998) sources and $z=1.7\pm0.1$ for the Conti \etal (1999) sources.\footnote{Two of the Jarvis \& MacAlpine (1998) sources have estimated redshifts of $z\approx 0.4$; omitting these sources gives $z=4.0\pm0.4$.} The soft-band upper limits correspond to average rest-frame 0.5--2.0~keV luminosities of $<4\times10^{41}$ erg s$^{-1}$ for the Jarvis \& MacAlpine (1998) sources and $<1\times10^{41}$ erg s$^{-1}$ for the Conti \etal (1999) sources. Therefore, any AGN activity must be intrinsically weak, in agreement with the low optical luminosities of the sources ($M_B<-20$, Jarvis \& MacAlpine 1998; $M_V<-17$, Conti \etal 1999; compare to Ho \etal 1997; 2001). Of course, the lack of X-ray emission is also consistent with no AGN activity in some, or all, of these objects. Significantly deeper \chandra\ observations ($\approx$ 5~Ms) are required to distinguish between these possibilities.

%
%%%%%%%%%%%%%%%%%%%%%%%%%%%%%%%%%%%%%%%%%%%%%%%%%%%%%%%%%%%%%%%%%%%%%%
\section{Discussion}
%%%%%%%%%%%%%%%%%%%%%%%%%%%%%%%%%%%%%%%%%%%%%%%%%%%%%%%%%%%%%%%%%%%%%%
%

The X-ray-to-optical flux ratios (see \S4.2) of the optically faint X-ray sources suggest the X-ray emission is due to AGN activity in the majority of cases. The red optical-to-near-IR colors (see \S4.1) suggest the majority of the optically faint X-ray sources are not normal BLAGN and reside in comparitively normal galaxies that are either dust extincted and/or lie at $z>1$. The flat X-ray spectral slopes (see \S4.3) and comparitively low incidence of X-ray variability (see \S4.4) further suggests that AGN activity is obscured in the majority of the sources. As some optically faint X-ray sources are among the brightest X-ray sources detected, an appreciable fraction of the optically faint X-ray sources could be luminous obscured QSOs.

Many X-ray background synthesis models predict a large number of luminous obscured QSOs (i.e.,\ $L_X>3\times10^{44}$~erg s$^{-1}$) at high redshift (e.g.,\ Wilman, Fabian, \& Nulsen 2000; Gilli \etal 2001), and obscured QSOs are expected within the unified model for AGN (e.g.,\ Antonucci 1993). However, the number of confirmed obscured QSOs in the local Universe is small (e.g.,\ Halpern \etal 1999; Franceschini \etal 2000; and references therein). At higher redshifts the situation is more promising as probable obscured QSOs have been detected in the \chandra\ Abell~1835 and Abell~2390 cluster fields and the \chandra\ Deep Field surveys (Crawford \etal 2000; Fabian \etal 2000; Paper I; Cowie \etal 2001; Norman \etal 2001). With the exception of the candidate obscured QSO in the \chandra\ Deep Field South (Norman \etal 2001) and some of the possible obscured QSOs in the cluster fields (Crawford \etal 2000), the candidate obscured QSOs are optically faint. Therefore, if a large number of obscured QSOs exist, many are likely be found within the optically faint X-ray source population. If the origin of the obscuration in these sources is absorption from gas and dust, they should produce powerful infrared emission and may contribute significantly to the infrared background radiation (e.g.,\ Puget \etal 1996; Schlegel \etal 1998).

In this section we review the redshifts of the most intensively studied optically faint X-ray sources to date, determine the redshift range of the optically faint X-ray sources, place constraints on the fraction of obscured QSOs within the optically faint X-ray population, and estimate their infrared fluxes.

\subsection{Redshifts of well-studied optically faint X-ray sources}

The redshifts of optically faint X-ray sources are almost completely unknown. To date, only three optically faint X-ray sources have optical spectroscopic redshifts (CDFS J033208.3-274153 at $z=2.453$; CXOHDFN J123633.5+621418 at $z=3.403$; CXOHDFN J123642.1+621332 at $z=4.424$); the first source is in the \chandra\ Deep Field South survey (Schreier \etal 2001) and the latter two sources are within our optically faint X-ray source sample (see \S3.1). The two optically faint X-ray sources with radio and sub-millimeter emission reported in \S5.1 have millimetric redshifts (CXOHDFN J123618.4+621551 has $z=$~$1.8^{+0.7}_{-0.6}$ and CXOHDFN J123646.1+621449 has $z=$~$2.3^{+0.8}_{-0.7}$; see Barger \etal 2000). There are also two optically faint X-ray sources with multi-band photometric redshifts: the $z=2.6^{+0.1}_{-0.2}$, $I=25.9$ (corrected for cluster lensing) source CXOU J215333.2+174211 (Cowie \etal 2001) and the $z\approx 2.7$, $I=25.8$ source CXOHDFN J123651.8+621221 (Dickinson 2000; Budav\'ari \etal 2000; Paper I).\footnote{There are two {\it ROSAT} Ultra-Deep Survey sources that fall just outside our definition of an optically faint source and have $23.5<I<24$; these sources have photometric redshifts of $z=1.9$ and $z=2.7$ (Lehmann \etal 2001).} The latter source is also within our optically faint X-ray sample. Both of these sources have been reported as obscured QSO candidates (Paper I; Cowie \etal 2001; Paper IV) although a photometric redshift fit for CXOHDFN J123651.8+621221 has never been published. Below we present our photometric redshift estimate of CXOHDFN J123651.8+621221 using the publicly available photometric redshift code {\sc hyperz} Version 1.1 (Bolzonella, Miralles, \& Pell\'o 2000).\footnote{This code is available at http://webast.ast.obs-mip.fr/hyperz/.} 

The optical-to-near-IR photometry for CXOHDFN J123651.8+621221 was provided by M.~Dickinson 2000, private communication. As CXOHDFN J123651.8+621221 is not in the Williams \etal (1996) catalog\footnote{This source lies close to a bright optical galaxy and was discovered in the radio (Richards \etal 1999) and near-IR wavebands (Dickinson \etal 2000) before it was identified at optical wavelengths.}; we determined the uncertainties in the magnitudes for each waveband from sources of similar magnitudes ($\pm0.1$ mags) in the Fern\'andez-Soto, Lanzetta, \& Yahil (1999) HDF-N source catalog. In performing the photometric redshift fitting, we used all the spectral templates provided with {\sc hyperz} and allowed up to 1 mag of visual extinction. The best photometric redshift solution is found to be a young ($\sim$1.4 Gyr old) elliptical galaxy at $z=2.75^{+0.13}_{-0.20}$ with 1 mag of visual extinction (see Figure~11); the uncertainities in the redshift determination correspond to the 90\% confidence level. The fit is best constrained by the 4000~\AA\ break, which falls between the $J$-band and $H$-band, and the Lyman limit, which enters the $F300W$ band and explains the non-detection of the source in that band. Similar fits in the same redshift range, although at a lower confidence level, were found for different spectral templates and different visual extinction constraints. Assuming $z=2.75$ and the column density of
%
%%%%%%%%%%%%%%%%%%%%%%%%%%%%%%%%%%%%%%%%%%%%%%%%%%%%%%%%%%%%%%%%%%%%%%
% 11 Photometric redshift determination for the QSOII source
%%%%%%%%%%%%%%%%%%%%%%%%%%%%%%%%%%%%%%%%%%%%%%%%%%%%%%%%%%%%%%%%%%%%%%
%
\vspace{0.2in}
\centerline{\includegraphics[angle=-90,width=9.0cm]{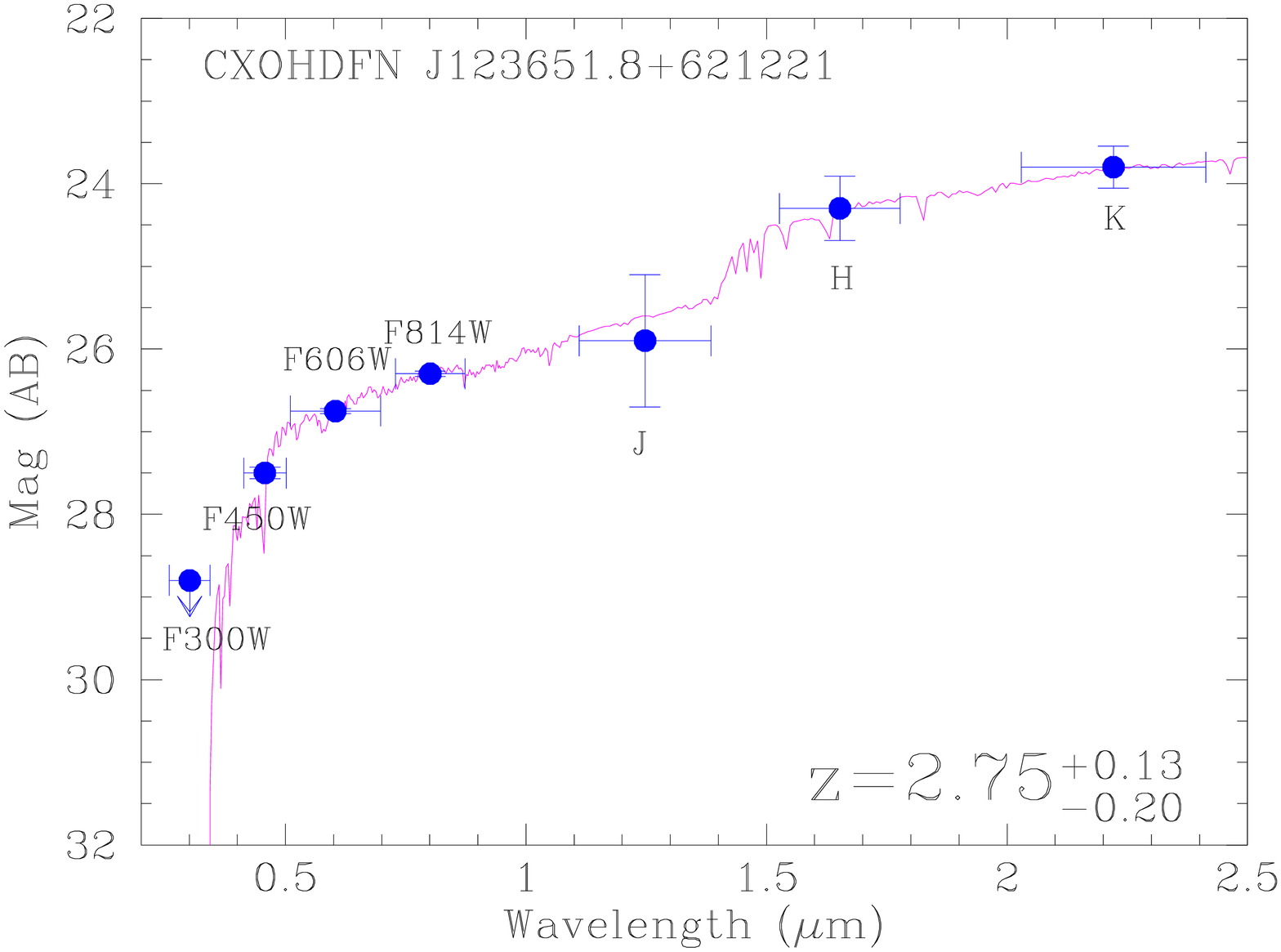}}
\vspace{0.1in}
\figcaption{The best-fitting photometric redshift solution for the HDF-N optically faint source CXOHDFN J123651.8+621221 found using {\sc hyperz} Version 1.1 (Bolzonella, Miralles, \& Pell\'o 2000). The solid curve shows the best-fit galaxy model, and the filled circles show the source photometry, (provided by M.~Dickinson 2000, private communication). The uncertainties in the band magnitudes are the average measured uncertainties for sources in the Fern\'andez-Soto, Lanzetta, \& Yahil (1999) HDF-N catalog with similar band magnitudes ($\pm$0.1 mag). The best-fit galaxy model is a young ($\sim$1.4 Gyr old) elliptical galaxy with 1 mag of extinction at $z=2.75^{+0.13}_{-0.20}$; the uncertainties in the redshift determination correspond to the 90\% confidence level.}
\label{fig:redshift1}
\vspace{0.2in}

%%%%%%%%%%%%%%%%%%%%%%%%%%%%%%%%%%%%%%%%%%%%%%%%%%%%%%%%%%%%%%%%%%%%%%

\noindent absorption given in Paper IV, CXOHDFN J123651.8+621221 would have an unabsorbed 0.5--8.0~keV rest frame luminosity of $\approx 3\times10^{44}$~erg s$^{-1}$, consistent with its obscured QSO status.

The redshift, $I$-band magnitude and X-ray luminosity of CXOHDFN J123651.8+621221 are similar to those of CXOU J215333.2+174211 (Cowie \etal 2001), suggesting they are very similar objects. By comparison CDFS J033208.3--274153 (Schreier \etal 2001), CXOHDFN J123642.1+621332 (Waddington \etal 1999; Paper IV), CXOHDFN J123618.4+621551 and CXOHDFN J123646.1+621449 are less luminous at X-ray energies and, unless their X-ray emission is Compton thick, are not obscured QSOs. CXOHDFN J123633.5+621418 is a BLAGN (see \S3.1), has comparitively blue colors ($I-K<3.0$) and a steep X-ray spectral slope typical of BLAGN (i.e.,\ $\Gamma=1.7$; see Table 2). Hence, because the majority of the optically faint X-ray sources have red optical-to-near-IR colors and/or flat X-ray spectral slopes, we do not believe that CXOHDFN J123633.5+621418 is a typical optically faint X-ray source. 

Based on this limited sample, two (29$^{+38}_{-18}$\%) of the seven optically faint X-ray sources with redshifts are probably obscured QSOs.

\subsection{Optically faint X-ray source redshift estimation}

Although we do not have sufficient photometric information to determine the redshifts of the optically faint X-ray sources on a source-by-source basis, we can estimate the probable 
%
%%%%%%%%%%%%%%%%%%%%%%%%%%%%%%%%%%%%%%%%%%%%%%%%%%%%%%%%%%%%%%%%%%%%%%
% 12 Comparison of redshifts to HDF-N photometrically determined redshifts
%%%%%%%%%%%%%%%%%%%%%%%%%%%%%%%%%%%%%%%%%%%%%%%%%%%%%%%%%%%%%%%%%%%%%%
%
\vspace{0.2in}
\centerline{\includegraphics[angle=0,width=9.0cm]{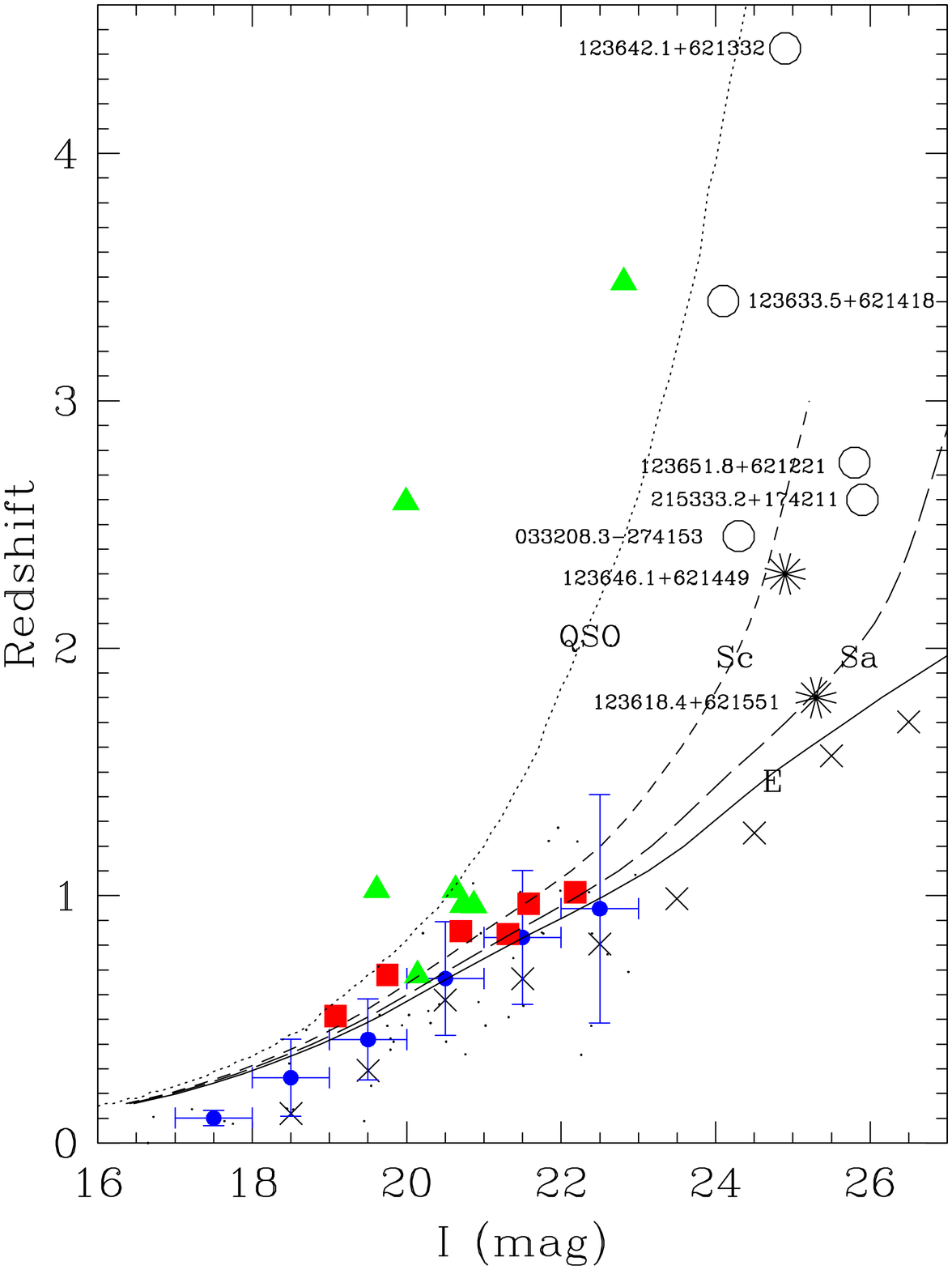}}
\vspace{0.1in}
\figcaption{\chandra\ source redshifts compared to the redshifts for different source types. The small dots are the \chandra\ sources with spectroscopic redshifts, the filled triangles are the BLAGN, the filled squares are the luminous NLAGN, the open circles are optically faint X-ray sources with spectroscopic or photometric redshifts (see \S6.1), and the stars are optically faint X-ray sources with millimetric redshifts (see \S5.1). The filled circles are the average spectroscopic redshifts for the $I<23$ \chandra\ sources; the width of each magnitude bin is shown as bars in the $x$-axis direction. The crosses are the average photometric redshifts for optical sources in the HDF-N (from Fern\'andez-Soto, Lanzetta, \& Yahil 1999). The solid, long-dashed and short-dashed curves are the redshift tracks of $M_{I}=-23$ E, Sa and Sc host galaxies. The dotted curve is the redshift track of an $M_{I}=-23$ QSO. The galactic $K$-corrections were taken from Poggianti (1997), and the QSO $K$-corrections were calculated with the QSO spectrum used in Figure~6. This figure suggests that if the optically faint X-ray sources are the high-redshift analogs of the optically bright X-ray sources, the majority should lie at $z\approx$~1--3; compare to Figure~7 of Barger \etal (2001a).}
\label{fig:redshift1}
\vspace{0.5in}

%%%%%%%%%%%%%%%%%%%%%%%%%%%%%%%%%%%%%%%%%%%%%%%%%%%%%%%%%%%%%%%%%%%%%%

\noindent redshift range from a comparison to the properties of the optically bright X-ray sources. In Figure~12 we have plotted the spectroscopic redshifts of the $I<23$ X-ray sources versus $I$-band magnitude (79\% of these sources have spectroscopic redshifts); for the non-BLAGN sources we have calculated the average redshift for each optical magnitude between $I=18$ and $I=23$. As a comparison to these data we have plotted the average photometric redshift for optical field galaxies for each optical magnitude between $I=18$ and $I=26$ using the Fern\'andez-Soto, Lanzetta, \& Yahil (1999) database of HDF-N photometric redshifts. We have also plotted the expected redshifts for $M_I=-23$ spiral and elliptical host galaxies and an $M_I=-23$ normal QSO.\footnote{$M_I<-23$ is equivalent to the classical QSO threshold of $M_B<-22.3$ (Schmidt \& Green 1983), adjusted to our assumed cosmology.}

Based on our analysis in \S4, we suggested that the majority of the optically faint X-ray sources are obscured AGN. The optical emission from an obscured AGN is dominated by the emission from the host galaxy and therefore has little bearing on the power of the X-ray source. However, as can be seen in Figure~12, the optically bright luminous NLAGN sources follow the track expected for an $M_I=-23$ host galaxy, showing that luminous obscured AGN reside in moderately luminous host galaxies. Assuming this trend continues to fainter optical magnitudes, the range in redshifts for the majority of the optically faint X-ray source population should be $z\approx$~1--3. This redshift range is in agreement with the red optical-to-near-IR colors of the majority of the optically faint X-ray sources (\S4.1) and the redshifts of the sources in \S6.1; this analysis is similar to that performed by Barger \etal (2001a), and quantitatively similar conclusions are reached. However, CXOHDFN J123642.1+621332 lies at a substantially higher redshift and shows that there can be exceptions.

Based on a simple hierarchial cold dark matter model and using constraints from the QSO X-ray luminosity function, Haiman \& Loeb (1999) predicted $\approx 15$ QSOs (i.e.,\ $L_X>10^{44}$ erg s$^{-1}$) at $z\simgt6$ at the depth and area of our survey. Any source at $z\simgt6$ would have extremely weak $I$-band emission due to Lyman-$\alpha$ leaving the $I$-band and consequently very red optical-to-near-IR colors (see Figure~6); an example of this is the $i'$-band drop-out source SDSSp J104433.04--012502.2 which lies at $z=5.8$ (Fan \etal 2000) and has $i^\prime-K=4.8$. We have 15 optically faint X-ray sources without $I$-band counterparts, exactly the number of $z\simgt6$ sources predicted by Haiman \& Loeb (1999). While none of these sources shows evidence for shorter wavelength counterparts in the Hogg \etal (2000) $U_n, G_r$, ${\cal R}$-band images or the Barger \etal (1999) $B$ and $V$-band images, these images are not sufficiently deep enough to provide strong constraints. However, it is unlikely that all of these sources lie at $z\simgt6$ based on three simple constraints. 
First, the $z\approx 2.75$ optically faint source in the HDF-N itself (CXOHDFN J123651.8+621221; see \S6.1) has $I=25.8$, 0.5 mags below the 2$\sigma$ $I$-band limit of the majority of our sources and would probably appear optically blank if it lay outside the HDF-N.\footnote{We also note that the optically faint source CXOU J215333.2+174211 (Cowie \etal 2001) in the Abell~2390 lensing cluster has $I=25.9$ and would also probably appear optically blank in our survey.} Although the statistics are limited, based on the area of the HDF-N itself, we would expect $\approx 13$ such sources within our whole field, very similar to the number of actual optically blank X-ray sources found. Second, the optically blank X-ray source CXOHDFN J123615.9+621516 appears to be associated with an optically faint X-ray source (see Figure~3 and \S3.1). Given the low surface density of optically faint X-ray sources the probability of a chance coincidence is extremely low ($\approx 0.1$\%) and therefore it is likely that the optically blank X-ray source lies at the same redshift as the optically faint X-ray source and hence at $z<6$. Third, K-S tests of the X-ray band ratio and full-band flux distributions give 83\% and 7\% probabilities, respectively, that the optically blank and optically faint distributions are consistent (see Figure~4 and Figure~9). The consistency between the X-ray band ratio distributions suggests that both source populations contain the same object types (i.e.,\ mostly obscured AGN) whilst the lower probability of consistency for the full-band flux distribution is probably due to the fact that the optically blank X-ray sources lie at fainter X-ray fluxes (see Figure~4). This evidence suggests that the majority of the optically blank X-ray sources are the extension of the optically faint X-ray source population to fainter $I$-band magnitudes.

\subsection{Limits on the number of obscured QSOs in the optically faint X-ray source population}

On the assumption that the majority of the optically faint X-ray sources are obscured AGN at $z=$~1--3, we can constrain the number of obscured QSOs by determining the minimum redshift for each optically faint X-ray source to produce a QSO luminosity in the X-ray band. For our analysis here, the adopted luminosity threshold for a QSO source is a rest-frame, full-band, unabsorbed X-ray luminosity of $>3\times10^{44}$ erg s$^{-1}$.

In \hbox{Figure~13} we show the minimum redshift distribution for all the optically faint X-ray sources to produce an unabsorbed QSO luminosity. The effect of absorption on the observed X-ray flux has been corrected for on a source-by-source basis assuming that the observed X-ray spectral slope is due to absorption of an underlying $\Gamma=2.0$ power law. The full range of minimum redshifts is broad ($1<z<10$); see also Table 1. From this estimation there can be 8 ($17^{+8}_{-6}$\%) obscured QSOs with $z\le3$ and two ($4^{+5}_{-3}$\%) obscured QSOs with $z\le2.0$ in the optically faint sample. However, this determination was made assuming the X-ray emission is obscured but seen directly. If some of these sources have Compton-thick absorption then the observed X-ray emission will be predominantly reflected, and the intrinsic X-ray luminosity could be at least an order of magnitude higher (e.g.,\ Bassani \etal 1999). Indeed, observations of local obscured AGN suggest that $\approx 50$\% of the sources have Compton-thick absorption (e.g.,\ Risaliti, Maiolino, \& Salvati 1999). Of the two most convincing obscured QSO candidates, IRAS 09104+4109 (Franceschini \etal 2000; Iwasawa, Fabian, \& Ettori 2001) almost certainly has Compton-thick absorption, and CDF-S 202 (Norman \etal 2001) possibly has Compton-thick absorption. In Figure~13 we also show the distribution of minimum redshifts for all the optically faint sources to produce a QSO X-ray luminosity assuming that each source is Compton thick. In this determination we have simply assumed that the scattering efficiency is 10\% (e.g.,\ Bassani \etal 1999) and that the scattered emission is not itself obscured. The full range of redshifts in this scenario is $1<z<6$. From this estimation 21 ($45^{+12}_{-10}$\%) of the sources could be obscured QSOs at $z\le3$ and 9 ($19^{+9}_{-6}$\%) could be obscured QSOs at $z\le2.0$. If $\approx$~50\% of the sources have Compton-thick absorption then the fraction of obscured QSOs will be somewhere between those given for the Compton-thin and Compton-thick cases. This is in reasonable agreement with our estimation based on a small sample of 7 optically faint X-ray sources with determined redshifts (i.e.,\ 29$^{+38}_{-18}$\%; see \S6.1).

Clearly a fraction of the optically faint X-ray source population are likely to be obscured QSOs. The population synthesis model of Gilli \etal (2001) predicts that obscured QSOs contribute $\approx 30$\% of the hard X-ray background. We do not find good agreement with this prediction as the optically faint X-ray source population only contributes $\approx $~21\% of the hard X-ray background and only a fraction of these sources are likely to be obscured QSOs. However, this 1 Ms observation has not fully resolved the hard X-ray background and the on-going optical spectroscopic identifications of the optically bright X-ray
%
%%%%%%%%%%%%%%%%%%%%%%%%%%%%%%%%%%%%%%%%%%%%%%%%%%%%%%%%%%%%%%%%%%%%%%
% 13 Minimum redshifts for sources to have QSO luminosities
%%%%%%%%%%%%%%%%%%%%%%%%%%%%%%%%%%%%%%%%%%%%%%%%%%%%%%%%%%%%%%%%%%%%%%
%
\vspace{0.2in}
\centerline{\includegraphics[angle=0,width=9.0cm]{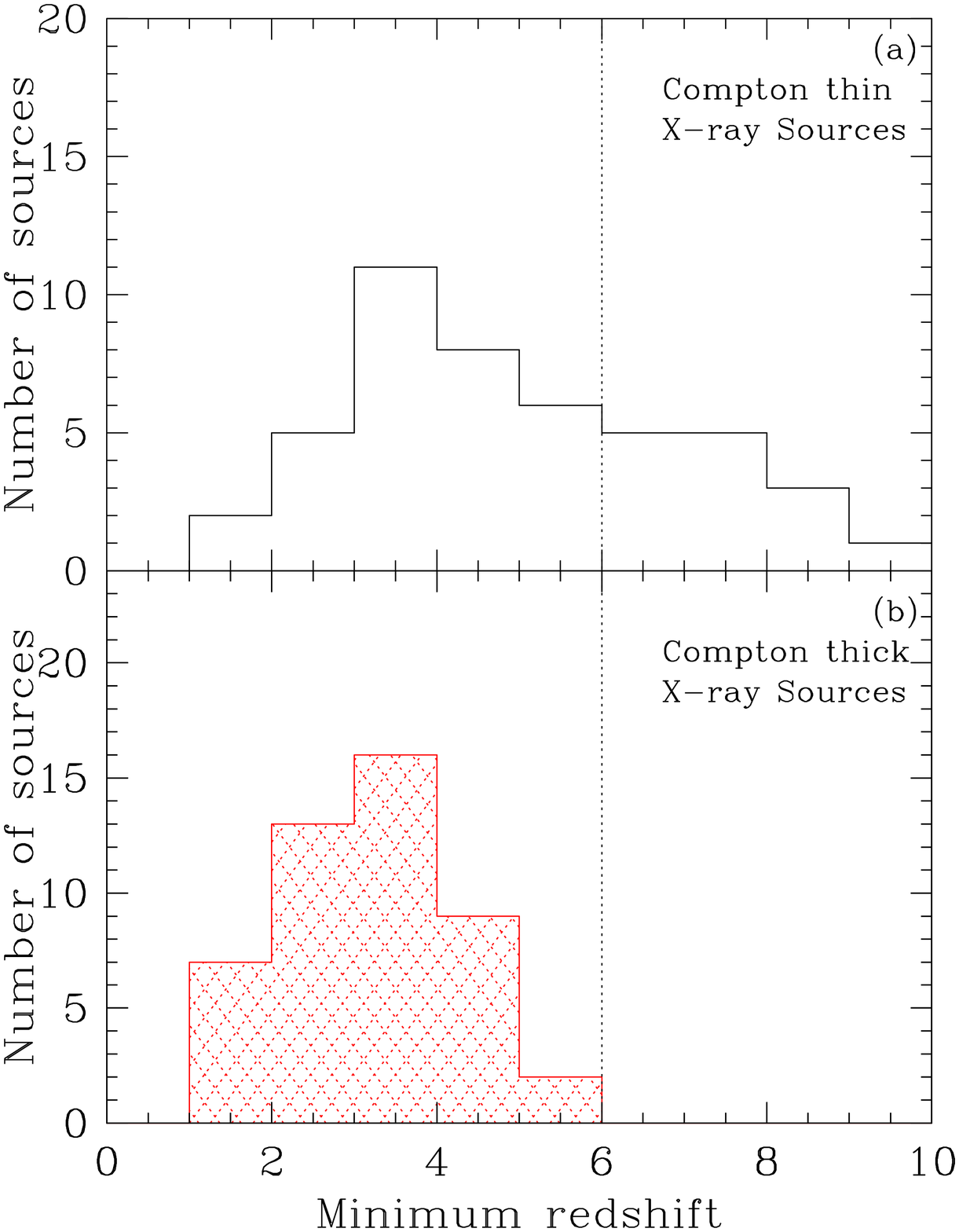}}
\vspace{0.1in}
\figcaption{The distribution of minimum redshifts for the optically faint X-ray sources required to produce a rest-frame, full-band, unabsorbed X-ray luminosity of $3\times10^{44}$ ergs s$^{-1}$ for (a) Compton thin sources and (b) Compton thick sources; a scattering efficiency of 10\% is assumed. The vertical dotted line shows the minimum redshift for an $I$-band drop-out source.}
\label{fig:qsoredshift}
\vspace{0.2in}

%%%%%%%%%%%%%%%%%%%%%%%%%%%%%%%%%%%%%%%%%%%%%%%%%%%%%%%%%%%%%%%%%%%%%%

\noindent sources may also reveal a number of obscured QSOs.

On the assumption that the optical magnitude implies the source redshift, the most promising obscured QSO candidates for $I= 24$--25 in our survey are those sources with hard band fluxes $>3\times 10^{-15}$~erg~cm$^{-2}$~s$^{-1}$ (see Figure~7b). Moderate depth, wide area surveys such as CHAMP (Wilkes \etal 2001) could therefore provide tighter constraints on the number of obscured QSOs in this optical magnitude range. Assuming all obscured QSOs have large X-ray to optical flux ratios, very deep optical observations (i.e.,\ $I=$~26--28) will be required to determine the optical counterparts of optically faint obscured QSOs at the flux limit of this survey.

\subsection{Infrared emission from optically faint X-ray sources}

According to the unified model for AGN (e.g.,\ Antonucci 1993), the origin of the absorption in obscured AGN is gas and dust within a circum-nuclear optically thick torus. In this model the ultraviolet and X-ray emission from the central AGN source heats the dust within the torus which re-emits this radiation in the infrared band (e.g.,\ Pier \& Krolik 1993; Granato \& Danese 1994; Efstathiou \& Rowan-Robinson 1995). If the optically faint X-ray sources are obscured AGN, they therefore should also produce powerful infrared emission. 

The only two optically faint X-ray sources (CXOHDFN J123642.1+621332 and CXOHDFN J123651.8+621221; see \S6.1) within the most sensitive ($\simlt~50~\mu$Jy) area of the deep {\it ISOCAM} HDF-N observation (Aussel \etal 1999) have faint 15~$\mu$m counterparts (e.g.,\ Waddington \etal 1999; Paper II; Paper IV). Eleven other optically faint X-ray sources lie within 2\amin\ of the center of the {\it ISOCAM} HDF-N field (see Figure~1); these sources are not detected but have upper limit fluxes ($<200~\mu$Jy) that are a factor of $\approx$~4 greater than the fluxes of the two detected sources. The hard X-ray (HX) to mid-IR (MIR) flux ratios of the detected sources (log${({{f_{\rm HX}}\over{f_{\rm MIR}}})}=$~--5 to --7) are consistent with that expected from AGN activity (e.g.,\ Alexander \etal 2001), suggesting that the infrared emission is produced by hot-dust emission within the dusty tori in these sources. 
The {\it ISOCAM} HDF-N observation is one of the three deepest 15~$\mu$m observations ever conducted; the other two deep 15~$\mu$m observations were taken in the Abell~2390 lensing cluster region (L\'emonon \etal 1998) and the HDF-S field (Oliver \etal 2001). The Abell~2390 lensing cluster region contains the optically faint X-ray source CXOU J215333.2+174211 (see \S6.1) which is detected at both 6.7~$\mu$m and 15~$\mu$m (Cowie \etal 2001); the infrared emission from this source is also consistent with that expected from a dusty torus (Wilman, Fabian, \& Gandhi 2000; Crawford \etal 2001). Assuming an average flux ratio of log${({{f_{\rm HX}}\over{f_{\rm MIR}}})}=$~--6, the optically faint X-ray sources should have 15~$\mu$m fluxes in the range $\approx$~10--450~$\mu$Jy. Sources with brighter X-ray fluxes or larger log${({{f_{\rm HX}}\over{f_{\rm MIR}}})}$ flux ratios would have 15~$\mu$m fluxes at the $>1$~mJy level. The combination of shallow, wide-area \chandra\ and \xmm\ surveys with infrared surveys, such as the European Large Area ISO Survey (Oliver \etal 2000) and the {\it SIRTF} (Fanson \etal 1998) First-Look Survey, could therefore be efficient ways of detecting X-ray bright obscured QSOs.\footnote{Details of the First-Look Survey can be found off the {\it SIRTF} home-page at http://sirtf.caltech.edu/.} To detect typical optically faint X-ray sources will require deeper infrared observations. Based on the sensitivity figures of Brandl (2000), a 2~ks {\it SIRTF} observation in the 24~$\mu$m {\it MIPS} band will detect a source at the 5$\sigma$ level with a 24~$\mu$m flux density of $\approx$~100~$\mu$Jy. A typical optically faint X-ray source should be detected at this level, assuming the spectral energy distribution (SED) of NGC~6240 (see below).

At longer wavelengths, the discovery of a ``Cosmic Far-Infrared Background'' between 140--240~$\mu$m (e.g.,\ Puget \etal 1996; Schlegel \etal 1998) has fueled great interest in the amount of dust-obscured activity in the Universe. While it is believed that a large fraction of this background emission is produced by star-forming galaxies (e.g.,\ Puget \etal 1999; Juvela \etal 2000; Scott \etal 2000), a non-negligible fraction may also be produced by AGN (e.g.,\ Almaini, Lawrence, \& Boyle 1999). As the SEDs of AGN and star-forming galaxies peak at $\approx$~60--100~$\mu$m, the $K$-correction for sources at redshifts of $z=$~1--3 is negative over the $\approx $~140--240~$\mu$m far-IR background band (e.g.,\ Blain \& Longair 1996; Puget \etal 1999). Therefore, if the majority of the optically faint X-ray source population lie at $z=$~1--3, they may contribute significantly to the far-IR background emission.

Although we have constraints on the mid-IR emission of optically faint X-ray sources, the production of the far-IR emission is not necessarily related to the AGN itself (e.g.,\ Alexander 2001). Indeed, the tight radio-to-far-IR correlation of galaxies (e.g.,\ Helou \etal 1985; Wunderlich \etal 1987) suggests that the far-IR emission in galaxies is produced by star-formation activity, even in many AGN sources. As the radio-to-far-IR correlation gives an estimate of the far-IR flux within the 40--120~$\mu$m band, we require an SED to determine the far-IR emission at other wavelengths. For our determination here we have chosen the SED of the luminous infrared galaxy NGC~6240; this galaxy shows evidence for both obscured AGN and star formation activity and has been used in other studies to determine the properties of sources detected in deep X-ray surveys (e.g.,\ Hasinger 2000; Barger \etal 2001a). The nuclear X-ray emission of NGC~6240 is obscured by Compton-thick material and, with a 2--10~keV luminosity $>10^{44}$ erg s$^{-1}$ (Iwasawa \& Comastri 1998; Vignati \etal 1999), it is a candidate obscured QSO. To determine how appropriate this galaxy is to optically faint X-ray sources, we have compared the SED of NGC 6240 to the multi-wavelength properties of CXOHDFN J123651.8+621221, our best-studied optically faint X-ray source (see Figure~14); the fluxes of NGC~6240 have been adjusted to show its emission at $z=2.75$. Although we cannot determine whether this SED is appropriate for the entire optically faint X-ray source population, the predicted sub-millimeter fluxes ($f_{850{\rm \mu m}}\approx$~0.5--2.0~mJy) are consistent with the low detection rate of bright sub-millimeter emission from X-ray sources (e.g.,\ Fabian \etal 2000; Paper I--II; Barger \etal 2001, 2001b). 

NGC 6240 and CXOHDFN J123651.8+621221 are clearly very similar, although CXOHDFN J123651.8+621221 is $\approx$~2--4 times more luminous in the X-ray and radio bands (see Figure~14). The differences in the X-ray emission at the lower energies are mainly due to absorption, as suggested by a comparison of the model fits to the X-ray emission of both sources (Vignati \etal 1999; \hbox{Paper IV}), and the X-ray spectral slopes are consistent at higher energies where the effect of absorption is less severe. The spectral slopes of the radio emission for both sources are $\alpha\sim$~0.7 (where $F_{\nu}\propto\nu^{-\alpha}$), the typical spectral slope for both normal galaxies and radio-quiet AGN. From the radio-to-far-IR correlation given in Barger \etal (2001a), we calculate a far-IR luminosity for CXOHDFN J123651.8+621221 of $3\times10^{46}$ erg s$^{-1}$, a factor of $\approx$~4 times greater than that of NGC~6240. Based on this far-IR luminosity, CXOHDFN J123651.8+621221 would be considered an Ultra-Luminous Infrared Galaxy (ULIRG; e.g.,\ Sanders \etal 1988; Genzel \etal 1998); see also Paper II. The deepest far-IR background source identification surveys to date have been performed with {\it ISOPHOT} in the 175~$\mu$m band. At the 5$\sigma$ sensitivity of these surveys ($f_{175{\rm \mu m}}\approx$~75--120~mJy; e.g.,\ Kawara \etal 1998; Puget \etal 1999; Juvela, Matilla, \& Lemke 2000), $\approx 10$\% of the far-IR background has been resolved. The estimated flux of CXOHDFN J123651.8+621221 at 175~$\mu$m is $\approx$~5 mJy, over an order of magnitude below the 5$\sigma$ sensitivity of the current {\it ISOPHOT} 175~$\mu$m surveys and possibly close to the resolution limit of the far-IR background (Puget \etal 1999).

The eight other optically faint X-ray sources with 1.4 GHz radio counterparts (see \S5.1) have either radio fluxes within a factor $\approx 3$ of that found for CXOHDFN J123651.8+621221 or clearly have a significant AGN contribution to their radio emission (e.g.,\ CXOHDFN J123642.1+621332; CXOHDFN J123707.2+621408; CXOHDFN J123721.2+621130); some of these sources may be brighter at far-infrared wavelengths than CXOHDFN J123651.8+621221 but should have 175~$\mu$m fluxes $<$~20~mJy. Assuming a redshift range of $z=$~1--3, the upper-limit far-IR luminosities for sources not detected with radio
%
%%%%%%%%%%%%%%%%%%%%%%%%%%%%%%%%%%%%%%%%%%%%%%%%%%%%%%%%%%%%%%%%%%%%%%
% 14 Comparison of SEDs
%%%%%%%%%%%%%%%%%%%%%%%%%%%%%%%%%%%%%%%%%%%%%%%%%%%%%%%%%%%%%%%%%%%%%%
%
\vspace{0.2in}
\centerline{\includegraphics[angle=-90,width=9.0cm]{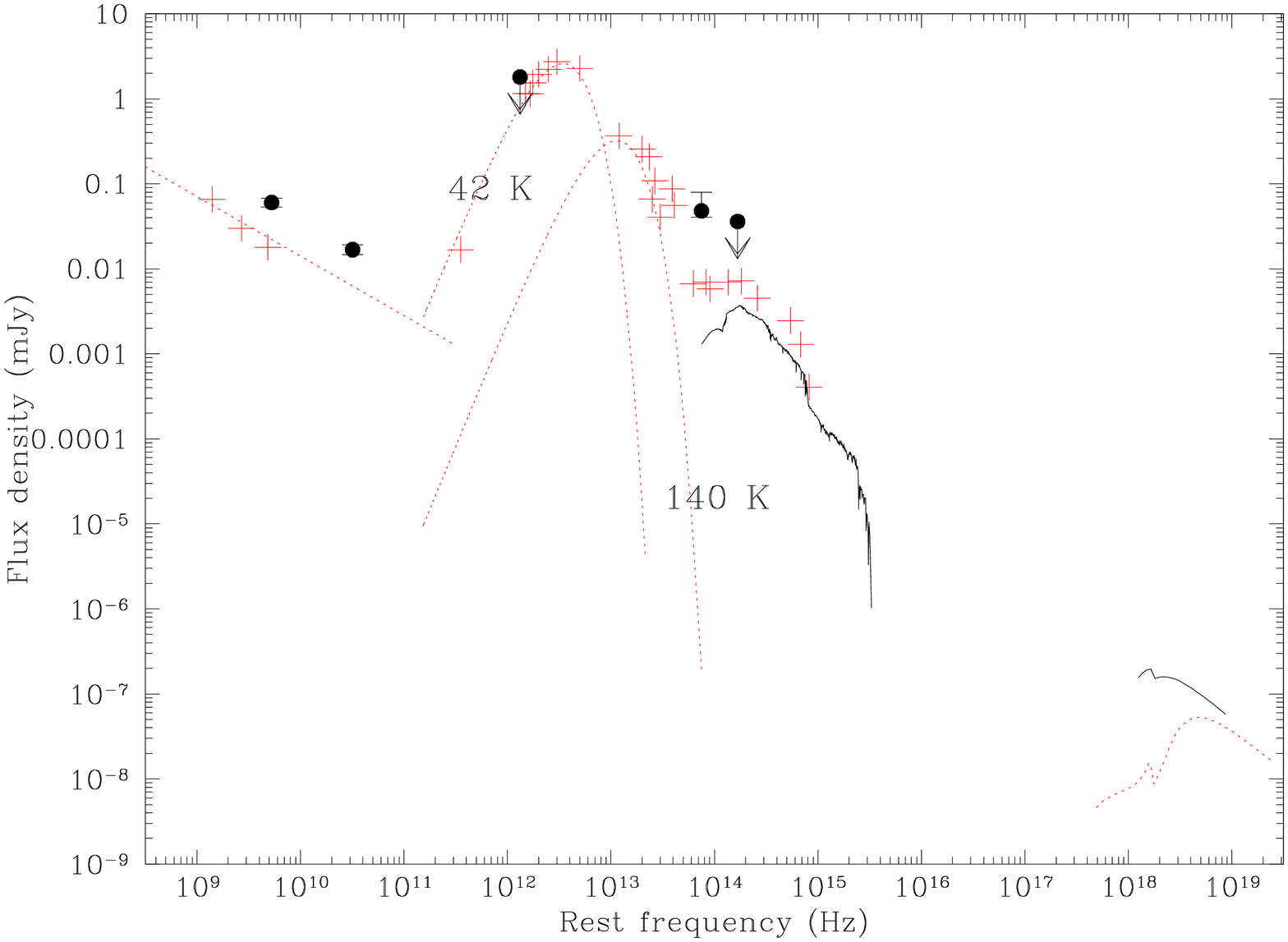}}
\vspace{0.1in}
\figcaption{A comparison between the multi-wavelength properties of CXOHDFN J123651.8+621221 and the SED of the luminous infrared galaxy NGC 6240. The filled circles show the data for CXOHDFN J123651.8+621221, and the crosses show the data for NGC 6240. The radio, sub-millimeter and infrared data for CXOHDFN J123651.8+621221 were taken from Richards (2000), Barger, Cowie, \& Richards (2000), and Aussel \etal (1999), respectively. The radio, sub-millimeter and infrared data for NGC 6240 were taken from Colbert \etal (1994), Lisenfeld \etal (2000), and Klass \etal (1997), respectively; the optical and near-IR data were taken from de Vaucouleurs \etal (1991) and Spinoglio \etal (1995) respectively. The solid curves show the photometric redshift model for CXOHDFN J123651.8+621221 (see Figure~11) in the optical-to-near-IR bands and the best fit absorbed power-law emission model for CXOHDFN J123651.8+621221 (Paper IV) in the X-ray band. The dotted lines show the power-law emission of NGC 6240 in the radio (Colbert \etal 1994), the modified black-body emission models (the dust temperatures are indicated) for NGC 6240 in the infrared band (Klass \etal 1997), and the best fit absorbed power-law emission model for NGC 6240 in the X-ray band (Vignati \etal 1999). The flux densities of the observations and models for NGC 6240 have been adjusted to $z=2.75$; therefore this figure shows the respective luminosities of NGC 6240 and CXOHDFN J123651.8+621221 over the full radio-to-X-ray wavelength range.}
\label{fig:redshift1}
\vspace{0.5in}

%%%%%%%%%%%%%%%%%%%%%%%%%%%%%%%%%%%%%%%%%%%%%%%%%%%%%%%%%%%%%%%%%%%%%%

\noindent emission range from $\approx$~(0.1--2.2)$\times10^{46}$ erg s$^{-1}$, and any source with $z\simgt$~1.6 could be a ULIRG. Following the same analysis as for CXOHDFN J123651.8+621221, the 175~$\mu$m fluxes are 6--3 mJy for $z=$~1--3. Therefore, on the assumption that the SED of NGC~6240 is appropriate for the other optically faint X-ray sources, it appears unlikely that any of the optically faint X-ray sources will be detected at the limit of the current {\it ISOPHOT} 175~$\mu$m surveys. We note, however, that any optically faint X-ray source with a bright sub-millimeter counterpart ($f_{850{\rm \mu m}}>$~3~mJy) could produce significant far-IR emission. For example, the two X-ray detected sub-millimeter sources reported in \S5.1 could have 175~$\mu$m fluxes of $\approx$~50~mJy.

The estimated flux level of the optically faint X-ray sources is within that achievable by {\it SIRTF} in the 160~$\mu$m {\it MIPS} band assuming no source confusion; however, in practice most observations will suffer source confusion before reaching these faint flux levels. Assuming the SED of NGC~6240, the estimated 70~$\mu$m fluxes of the optically faint X-ray sources range from 0.5--2.0 mJy. Based on the sensitivity predictions of Brandl (2000), these sources should be detectable by {\it SIRTF} at the 5$\sigma$ level in the 70~$\mu$m {\it MIPS} band with $>3.5$~ks exposures.

%
%%%%%%%%%%%%%%%%%%%%%%%%%%%%%%%%%%%%%%%%%%%%%%%%%%%%%%%%%%%%%%%%%%%%%%
\section{Conclusions}
%%%%%%%%%%%%%%%%%%%%%%%%%%%%%%%%%%%%%%%%%%%%%%%%%%%%%%%%%%%%%%%%%%%%%%
%

We have used a 1~Ms \chandra\ exposure of the Hubble Deep Field North (HDF-N) region and $8.4\amin\times8.4\amin$ area within the Hawaii flanking-field region to provide constraints on the nature of optically faint ($I\ge24$) X-ray sources. Our main results are the following:

(i)~We have detected 47 (33\% of all the X-ray sources in this survey; a source density of $\approx 2,400^{+400}_{-350}$ deg$^{-2}$) optically faint X-ray sources. These sources contribute $\approx 14$\% of the X-ray background in the soft band and $\approx 21$\% of the X-ray background in the hard band. The fraction of optically faint sources within the X-ray source population appears to be approximately constant (at $\approx 35$\%) for full-band fluxes between $3\times10^{-14}$~erg~cm$^{-2}$~s$^{-1}$ and $\approx 2\times10^{-16}$~erg~cm$^{-2}$~s$^{-1}$. See \S3.

(ii)~The large X-ray-to-optical flux ratios, red optical-to-near-IR colors, flat X-ray spectral slopes, and X-ray variability properties of the optically faint X-ray sources suggest that obscured AGN activity is present in the majority of cases. Assuming the optically faint X-ray source population is the high-redshift analog of the optically bright X-ray source population, the majority of the optically faint X-ray source population should lie at $z=$~1--3. From these results we calculate that a significant fraction ($\approx$~5--45\%) of optically faint X-ray sources could be obscured QSOs (rest-frame unabsorbed 0.5--8.0~keV luminosity $>3\times10^{44}$~erg~s$^{-1}$) at $z\le3$; from the analysis of a small sample of 7 optically faint X-ray sources with redshifts, two (29$^{+38}_{-18}$\%) are probably obscured QSOs. All but $\approx$~15 of the optically faint X-ray sources have 2$\sigma$ $I$-band counterparts, and hence there are unlikely to be more than $\approx$~15 sources at $z>6$. We provide evidence that the true number of $z>6$ sources is likely to be considerably lower. There are unlikely to be many optically faint Galactic sources or clusters of galaxies within our sample of optically faint X-ray sources. See \S4 and \S6.1--6.3.

(iii)~We determine the photometric redshift of one source, CXOHDFN J123651.8+621221, with seven band photometry to be $z=2.75^{+0.13}_{-0.20}$. We find the radio-to-X-ray properties of this source to be similar to those of the luminous infrared galaxy NGC~6240, although CXOHDFN J123651.8+621221 is $\approx$~2--4 times more luminous in the X-ray and radio bands. Based purely on its calculated far-IR luminosity, CXOHDFN J123651.8+621221 would be considered a ULIRG. See \S6.1 and \S6.4.

(iv)~We estimate that the vast majority of the optically faint X-ray sources have faint 175~$\mu$m ($\approx$~3--6 mJy) counterparts; however, sources with bright sub-millimeter counterparts (i.e.,\ $f_{850{\rm \mu m}}>$~3~mJy) could have substantially brighter 175~$\mu$m fluxes. Therefore, the estimated 175~$\mu$m fluxes of a typical optically faint X-ray source will be approximately an order of magnitude below that achieved by the current 175~$\mu$m {\it ISOPHOT} surveys. Hence these sources are unlikely to contribute significantly to the far-IR (140--240~$\mu$m) background radiation. However, the only two optically faint X-ray sources within the most sensitive region of the {\it ISOCAM} HDF-N survey do have faint ($\simlt$~50~$\mu$Jy) counterparts at 15~$\mu$m; the hard-band X-ray to mid-IR flux ratios of these sources are consistent with that expected from an AGN source. These results suggest moderate-to-deep 24~$\mu$m and 70~$\mu$m {\it SIRTF} observations should detect a large number of optically faint obscured QSO sources. X-ray observations will provide the most direct determination of obscured QSO activity. See \S6.4.

(v)~Nine of the optically faint X-ray sources have $\mu$Jy radio source counterparts; this is $\approx 53^{+24}_{-17}$\% of the optically faint $\mu$Jy radio source sample in our area. The nature of the X-ray emission from the majority of these detected sources is clearly obscured AGN activity. Two sources are also detected at sub-millimeter wavelengths. The nature of the X-ray emission in these sources could be luminous star formation activity. A stacking analysis of the X-ray undetected $\mu$Jy radio sources yields a possible detection (at $98.3$\% confidence) in the soft band. This emission may be produced by star formation activity from Arp~220-like sources at $z\approx 1.5$ or NGC~3256-like sources at $z\approx 3.0$. See \S5.1.

(vi)~None of the optically selected AGN candidates in the HDF-N itself has been detected either individually or with a stacking analysis. This suggests that these sources have low X-ray luminosities, in general agreement with their absolute optical magnitudes. Significantly deeper \chandra\ observations ($\approx 5$~Ms) are required to uncover any normal AGN activity within these sources. See \S5.2.

%
%%%%%%%%%%%%%%%%%%%%%%%%%%%%%%%%%%%%%%%%%%%%%%%%%%%%%%%%%%%%%%%%%%%%%%
\section*{Acknowledgements}
%%%%%%%%%%%%%%%%%%%%%%%%%%%%%%%%%%%%%%%%%%%%%%%%%%%%%%%%%%%%%%%%%%%%%%
%

This work would not have been possible without the support of the entire \chandra\ and ACIS teams; we particularly thank P.~Broos and L.~Townsley for data analysis software and CTI correction support.
We thank E.~Feigelson, Z.~Haiman, P.~Lira, G.~Pavlov, G.~Richards and C.~Vignali for helpful discussions and to the anonymous referee for useful comments that improved the presentation of the paper.
We are grateful to M.~Dickinson for providing photometric data and to M.~Bolzonella, J.-M.~Miralles and R.~Pello for making {\sc hyperz} available. We thank A.~Barger, L.~Cowie, D.~Hogg and C.~Steidel for making their optical and near-IR images publicly available.
We acknowledge the financial support of
NASA grant NAS~8-38252 (GPG, PI),
NSF CAREER award AST-9983783 (DMA, WNB, FEB),  
NASA GSRP grant NGT5-50247 and 
the Pennsylvania Space Grant Consortium (AEH), and
NSF grant AST-9900703~(DPS).

%
%%%%%%%%%%%%%%%%%%%%%%%%%%%%%%%%%%%%%%%%%%%%%%%%%%%%%%%%%%%%%%%%%%%%%%

%
%%%%%%%%%%%%%%%%%%%%%%%%%%%%%%%%%%%%%%%%%%%%%%%%%%%%%%%%%%%%%%%%%%%%%%%%%%%%%%%%%%%%%%
% TABLES
%%%%%%%%%%%%%%%%%%%%%%%%%%%%%%%%%%%%%%%%%%%%%%%%%%%%%%%%%%%%%%%%%%%%%%%%%%%%%%%%%%
%

\clearpage

%
%%%%%%%%%%%%%%%%%%%%%%%%%%%%%%%%%%%%%%%%%%%%%%%%%%%%%%%%%%%%%%%%%%%%%%
% TABLE 1: x-ray optically faint table
%%%%%%%%%%%%%%%%%%%%%%%%%%%%%%%%%%%%%%%%%%%%%%%%%%%%%%%%%%%%%%%%%%%%%%
%

\renewcommand{\baselinestretch}{1.2}

%
%%%%%%%%%%%%%%%%%%%%%%%%%%%%%%%%%%%%%%%%%%%%%%%%%%%%%%%%%%%%%%%%%%%%%%
% TABLE 1
%%%%%%%%%%%%%%%%%%%%%%%%%%%%%%%%%%%%%%%%%%%%%%%%%%%%%%%%%%%%%%%%%%%%%%
%

\begin{deluxetable}{llcccccccrrrrrr}
%\rotate
%\tablecolumns{14}
\tabletypesize{\tiny}
\tablewidth{0pt}
\tablecaption{Optically faint X-ray source properties}
\tablehead{
\multicolumn{2}{c}{Coordinates} &
\colhead{}                  &
\colhead{}                  &
\multicolumn{3}{c}{Counts}      &
\colhead{Band}                  &
\colhead{}                  &
\multicolumn{3}{c}{Flux}        &
\multicolumn{2}{c}{QSO redshift}  \\
\colhead{$\alpha_{2000}$}       &
\colhead{$\delta_{2000}$}       &
\colhead{$I^{\rm a}$}           &
\colhead{$I-K^{\rm b}$}         &
\colhead{FB$^{\rm c}$}          &
\colhead{SB$^{\rm c}$}          &
\colhead{HB$^{\rm c}$}          &
\colhead{Ratio$^{\rm d}$}       &
\colhead{$\Gamma$$^{\rm e}$}    &
\colhead{FB$^{\rm f}$}          &
\colhead{SB$^{\rm f}$}          &
\colhead{HB$^{\rm f}$}          &
\colhead{C.-thin$^{\rm g}$}     &
\colhead{C.-thick$^{\rm h}$}}
\startdata
12 36 13.07&                   +62 12 24.1&                24.7&              $<$3.6&                      74.2$\pm$11.1&                       23.0$\pm$6.7&                       53.9$\pm$9.8&              $2.39^{+1.08}_{-0.69}$&                0.01&      2.04&      0.13&      2.03&    $>$3.1&    $>$2.3\\
12 36 14.11&                   +62 10 17.4&             $>$25.3&                 $-$&                       25.4$\pm$9.2&                        7.3$\pm$5.8&                       15.2$\pm$7.9&              $2.12^{+8.41}_{-1.44}$&                0.10&      0.59&      0.04&      0.49&    $>$4.9&    $>$3.6\\
12 36 14.45&                   +62 10 45.6&                24.2&              $<$3.1&                      84.2$\pm$12.2&                       18.7$\pm$6.7&                      66.7$\pm$11.0&              $3.59^{+2.10}_{-1.12}$&               -0.28&      2.30&      0.09&      2.32&    $>$2.7&    $>$2.2\\
12 36 15.89&                   +62 15 15.5&             $>$25.3&              $>$4.2&                      81.5$\pm$18.7&                            $<$20.7&                      71.9$\pm$20.7&                             $>$3.52&            $<$-0.26&      2.19&   $<$0.10&      2.48&    $>$2.7&    $>$2.3\\
12 36 16.03&                   +62 11 07.9&                24.6&                 3.9&                     656.8$\pm$28.6&                     218.2$\pm$16.5&                     439.6$\pm$23.7&              $2.03^{+0.20}_{-0.18}$&                0.13&     14.94&      1.07&     13.99&    $>$1.3&    $>$1.0\\
12 36 16.11&                   +62 15 13.5&                24.3&                 2.8&                      95.6$\pm$21.2&                      25.9$\pm$17.2&                      43.4$\pm$14.0&              $1.70^{+3.40}_{-0.87}$&                0.28&      1.99&      0.13&      1.32&    $>$3.0&    $>$2.1\\
12 36 19.18&                   +62 14 41.6&                24.4&              $<$3.3&                     335.6$\pm$20.8&                     239.3$\pm$17.2&                      93.7$\pm$11.9&              $0.40^{+0.06}_{-0.06}$&                1.64&      3.31&      1.19&      2.08&    $>$2.7&    $>$1.3\\
12 36 20.52&                   +62 12 38.1&                24.7&                 4.6&                       29.4$\pm$7.9&                       16.3$\pm$5.8&                       15.0$\pm$6.3&              $0.93^{+0.65}_{-0.46}$&                0.88&      0.49&      0.09&      0.44&    $>$5.7&    $>$3.4\\
12 36 21.11&                   +62 13 03.6&                25.2&              $<$4.1&                       19.2$\pm$6.8&                        9.1$\pm$4.8&                            $<$11.9&                             $<$1.33&             1.40&      0.30&      0.06&   $<$0.38&    $>$6.2&    $>$4.1\\
12 36 21.93&                   +62 16 03.2&                25.5&              $<$4.4&                       34.8$\pm$8.6&                            $<$12.7&                       27.0$\pm$7.7&                             $>$2.15&             $<$0.09&      0.78&   $<$0.06&      0.84&    $>$4.3&    $>$3.2\\
12 36 22.66&                   +62 10 28.5&                24.7&                 3.9&                     418.1$\pm$23.1&                     289.0$\pm$18.7&                     120.9$\pm$13.4&              $0.42^{+0.06}_{-0.05}$&                1.59&      4.39&      1.48&      2.80&    $>$2.4&    $>$1.2\\
12 36 23.69&                   +62 10 09.0&             $>$25.3&                 $-$&                       33.8$\pm$9.0&                       17.9$\pm$6.5&                       14.9$\pm$7.0&              $0.84^{+0.62}_{-0.45}$&                0.99&      0.49&      0.09&      0.38&    $>$5.5&    $>$3.2\\
12 36 27.28&                   +62 13 08.3&             $>$25.3&                 $-$&                       10.0$\pm$5.6&                            $<$10.3&                        6.0$\pm$4.9&                             $>$0.59&             1.40&      0.11&   $<$0.05&      0.14&   $>$10.0&    $>$5.3\\
12 36 27.54&                   +62 12 18.2&             $>$25.3&                 $-$&                       28.3$\pm$7.4&                       20.4$\pm$6.1&                        9.9$\pm$5.6&              $0.49^{+0.34}_{-0.30}$&                1.48&      0.31&      0.10&      0.23&    $>$7.2&    $>$3.5\\
12 36 31.27&                   +62 09 58.3&                24.9&              $<$3.8&                       18.6$\pm$7.4&                       17.8$\pm$6.3&                            $<$11.0&                             $<$0.62&             1.40&      0.22&      0.09&   $<$0.26&    $>$8.0&    $>$4.1\\
12 36 33.49&                   +62 14 18.1&                24.1&              $<$3.0&                     110.7$\pm$12.4&                      86.6$\pm$10.7&                       28.8$\pm$7.2&              $0.34^{+0.10}_{-0.09}$&                1.74&      1.00&      0.42&      0.61&    $>$4.5&    $>$2.0\\
12 36 33.76&                   +62 13 13.8&                24.7&                 4.3&                       40.3$\pm$8.3&                             $<$9.8&                       35.6$\pm$7.9&                             $>$3.67&            $<$-0.29&      1.06&   $<$0.05&      1.19&    $>$3.6&    $>$3.0\\
12 36 33.86&                   +62 13 27.7&                24.1&              $<$3.0&                             $<$9.7&                        5.8$\pm$4.2&                             $<$4.8&                             $<$0.84&             1.40&   $<$0.11&      0.03&   $<$0.11&    $>$9.9&    $>$5.4\\
12 36 34.47&                   +62 09 41.1&             $>$25.3&                 $-$&                       25.8$\pm$8.1&                       13.3$\pm$5.8&                       11.1$\pm$6.4&              $0.84^{+0.81}_{-0.55}$&                1.40&      0.31&      0.07&      0.28&    $>$6.2&    $>$3.6\\
12 36 35.27&                   +62 11 51.7&                24.5&                 4.1&                       23.9$\pm$7.0&                        9.9$\pm$4.8&                       12.7$\pm$5.8&              $1.29^{+1.37}_{-0.73}$&                1.40&      0.29&      0.05&      0.32&    $>$5.7&    $>$3.7\\
12 36 36.89&         +62 13 20.2$^{\rm i}$&             $>$25.3&              $>$3.4&                       18.2$\pm$6.3&                        7.6$\pm$4.4&                       11.9$\pm$5.6&              $1.58^{+2.33}_{-0.94}$&                1.40&      0.20&      0.04&      0.27&    $>$6.1&    $>$4.1\\
12 36 38.97&                   +62 10 41.4&             $>$25.3&                 $-$&                       51.4$\pm$9.4&                       33.7$\pm$7.5&                       17.2$\pm$6.6&              $0.51^{+0.24}_{-0.22}$&                1.44&      0.56&      0.16&      0.39&    $>$5.4&    $>$2.7\\
12 36 39.65&                   +62 09 36.7&             $>$25.3&              $>$4.5&                       14.3$\pm$6.9&                        9.5$\pm$5.3&                            $<$12.1&                             $<$1.29&             1.40&      0.19&      0.06&   $<$0.33&    $>$7.2&    $>$4.6\\
12 36 42.11&         +62 13 31.6$^{\rm i}$&      24.9$^{\rm j}$&       3.5$^{\rm j}$&                       22.7$\pm$6.8&                       20.4$\pm$6.0&                             $<$9.5&                             $<$0.47&             1.40&      0.25&      0.10&   $<$0.21&    $>$8.0&    $>$3.8\\
12 36 46.05&         +62 14 49.0$^{\rm i}$&                24.9&              $<$3.8&                       12.6$\pm$5.7&                        9.2$\pm$4.6&                             $<$8.8&                             $<$0.98&             1.40&      0.14&      0.04&   $<$0.20&    $>$8.3&    $>$4.8\\
12 36 47.95&                   +62 10 19.9&             $>$25.3&              $>$4.6&                       36.2$\pm$8.3&                       11.6$\pm$5.3&                       24.4$\pm$7.2&              $2.11^{+1.87}_{-0.91}$&                0.10&      0.96&      0.07&      0.89&    $>$4.2&    $>$3.1\\
12 36 48.30&                   +62 14 56.3&             $>$25.3&                 $-$&                      66.4$\pm$10.1&                       30.4$\pm$6.9&                       36.1$\pm$7.9&              $1.21^{+0.44}_{-0.35}$&                0.61&      1.12&      0.14&      0.98&    $>$3.8&    $>$2.4\\
12 36 51.75&         +62 12 21.4$^{\rm i}$&      25.8$^{\rm j}$&       3.7$^{\rm j}$&                     148.1$\pm$14.2&                       60.1$\pm$9.2&                      91.9$\pm$11.6&              $1.54^{+0.34}_{-0.28}$&                0.37&      2.81&      0.28&      2.59&    $>$2.6&    $>$1.8\\
12 36 51.83&                   +62 15 04.9&                24.5&                 3.5&                     165.7$\pm$14.9&                     110.5$\pm$12.0&                       56.3$\pm$9.3&              $0.52^{+0.11}_{-0.10}$&                1.43&      2.13&      0.63&      1.52&    $>$3.3&    $>$1.7\\
12 36 54.57&                   +62 11 11.0&                25.3&                 4.2&                       30.7$\pm$7.7&                       12.1$\pm$5.2&                       20.6$\pm$6.7&              $1.70^{+1.38}_{-0.75}$&                0.28&      0.77&      0.07&      0.75&    $>$4.8&    $>$3.3\\
12 36 56.58&                   +62 15 13.2&             $>$25.3&                 $-$&                       24.3$\pm$6.9&                             $<$7.7&                       23.2$\pm$6.8&                             $>$3.09&            $<$-0.17&      0.60&   $<$0.04&      0.75&    $>$4.6&    $>$3.7\\
12 36 58.83&                   +62 10 22.3&                24.1&                 4.0&                       42.6$\pm$8.9&                            $<$10.3&                       39.9$\pm$8.6&                             $>$3.87&            $<$-0.34&      1.20&   $<$0.05&      1.41&    $>$3.5&    $>$2.9\\
12 37 00.46&                   +62 15 08.9&                24.1&                 3.7&                      99.1$\pm$11.9&                       60.3$\pm$9.3&                       43.9$\pm$8.6&              $0.74^{+0.20}_{-0.17}$&                1.11&      1.66&      0.37&      1.38&    $>$3.7&    $>$2.1\\
12 37 02.60&                   +62 12 44.0&                24.5&                 3.9&                       32.1$\pm$7.7&                       22.8$\pm$6.3&                            $<$10.8&                             $<$0.48&             $>$1.40&      0.35&      0.11&   $<$0.24&    $>$6.8&    $>$3.3\\
12 37 02.82&                   +62 16 01.3&                24.9&                 3.6&                     193.4$\pm$16.1&                     134.7$\pm$13.2&                      65.8$\pm$10.2&              $0.49^{+0.09}_{-0.09}$&                1.48&      2.10&      0.67&      1.51&    $>$3.2&    $>$1.6\\
12 37 04.08&                   +62 11 55.2&                24.5&              $<$3.4&                       14.3$\pm$6.0&                             $<$9.7&                       11.2$\pm$5.6&                             $>$1.16&             1.40&      0.16&   $<$0.05&      0.26&    $>$7.4&    $>$4.6\\
12 37 04.86&                   +62 16 01.6&                25.0&                 3.9&                     647.9$\pm$28.1&                     416.0$\pm$22.1&                     249.6$\pm$18.0&              $0.60^{+0.06}_{-0.05}$&                1.30&      7.77&      2.04&      5.94&    $>$1.8&    $>$1.0\\
12 37 05.10&                   +62 16 34.8&             $>$25.3&                 $-$&                      87.2$\pm$11.7&                       43.9$\pm$8.3&                       42.8$\pm$8.8&              $0.98^{+0.30}_{-0.25}$&                0.83&      1.32&      0.21&      1.10&    $>$3.6&    $>$2.2\\
12 37 07.23&         +62 14 08.0$^{\rm i}$&                25.0&                 5.0&                       54.7$\pm$9.3&                       24.2$\pm$6.4&                       32.2$\pm$7.7&              $1.35^{+0.58}_{-0.43}$&                0.50&      1.00&      0.12&      0.92&    $>$4.0&    $>$2.6\\
12 37 11.97&         +62 13 25.2$^{\rm i}$&                25.0&              $<$3.9&                       22.2$\pm$6.9&                             $<$9.7&                       20.4$\pm$6.7&                             $>$2.14&             $<$0.09&      0.56&   $<$0.05&      0.72&    $>$5.2&    $>$3.8\\
12 37 12.11&                   +62 12 11.3&                25.9&              $<$4.8&                       15.2$\pm$6.3&                             $<$9.7&                       10.3$\pm$5.7&                             $>$1.07&             1.40&      0.19&   $<$0.05&      0.27&    $>$7.3&    $>$4.5\\
12 37 12.66&                   +62 13 42.5&             $>$25.3&                 $-$&                       13.7$\pm$5.9&                             $<$8.5&                            $<$11.4&                                 $-$&                1.40&      0.16&   $<$0.04&   $<$0.28&    $>$7.2&    $>$4.7\\
12 37 13.70&                   +62 15 45.7&             $>$25.3&              $>$3.8&                       19.8$\pm$7.1&                             $<$9.1&                       17.7$\pm$6.7&                             $>$1.94&             $<$0.16&      0.42&   $<$0.04&      0.53&    $>$5.6&    $>$4.0\\
12 37 13.89&                   +62 14 58.0&                25.7&              $<$4.6&                       27.6$\pm$7.6&                            $<$12.0&                       24.4$\pm$7.1&                             $>$2.04&             $<$0.12&      0.64&   $<$0.06&      0.80&    $>$4.8&    $>$3.5\\
12 37 16.51&                   +62 16 43.2&                24.7&              $<$3.6&                      33.7$\pm$16.5&                        6.6$\pm$9.4&                      25.7$\pm$14.1&             $3.93^{+13.39}_{-3.16}$&               -0.35&      0.97&      0.03&      0.92&    $>$3.9&    $>$3.2\\
12 37 19.03&                   +62 10 25.6&                25.9&              $<$4.8&                      66.9$\pm$11.2&                       53.3$\pm$9.2&                       13.9$\pm$7.2&              $0.26^{+0.15}_{-0.14}$&                1.88&      0.60&      0.28&      0.31&    $>$6.1&    $>$2.4\\
12 37 22.72&                   +62 09 35.2&                24.1&              $<$3.0&                     111.8$\pm$14.1&                      59.6$\pm$10.0&                      50.0$\pm$10.7&              $0.85^{+0.25}_{-0.22}$&                0.97&      1.68&      0.30&      1.34&    $>$3.3&    $>$2.0\\
\enddata

\tablenotetext{a}{$I$-band Vega-based magnitude.}
\tablenotetext{b}{Calculated $I-K$ color determined from $I-(HK^\prime-0.3)$ following Barger et~al. (1999).}
\tablenotetext{c}{Source counts and errors taken from Paper V. ``FB'' indicates full band, ``SB'' indicates soft band, and ``HB'' indicates hard band.}
\tablenotetext{d}{Ratio of the counts between the 2.0--8.0~keV and 0.5--2.0~keV bands. The errors were calculated following the ``numerical method' described in \S1.7.3 of Lyons (1991). Taken from Paper V.}
\tablenotetext{e}{Photon index for the 0.5--8.0~keV band, calculated from the band ratio. Taken from Paper V. The photon index for those sources with a low number of counts have been set to $\Gamma =1.4$, a value representative of the X-ray background spectral slope; see Paper V.}
\tablenotetext{f}{Fluxes are in units of $10^{-15}$~erg~cm$^{-2}$~s$^{-1}$. These fluxes have been taken from Paper V. They have not been corrected for Galactic absorption. ``FB'' indicates full band, ``SB'' indicates soft band, and ``HB'' indicates hard band.}
\tablenotetext{g}{Minimum redshift for a source to have a rest-frame unabsorbed 0.5--8.0~keV luminosity $>3\times10^{44}$~ergs~s$^{-1}$. The unobscured X-ray emission is assumed to be a $\Gamma=2.0$ power law. See \S6.3 for details.}
\tablenotetext{h}{Minimum redshift for a source to have a rest-frame unabsorbed 0.5--8.0~keV luminosity $>3\times10^{44}$~ergs~s$^{-1}$. The source is assumed to be Compton-thick, and the unobscured X-ray emission is assumed to be a $\Gamma=2.0$ power law. See \S6.3 for details.}
\tablenotetext{i}{Optically faint $\mu$Jy radio source (Richards et~al. 1999); see Table 2.}
\tablenotetext{j}{$I$-band magnitude determined with the WFPC2 $I814W$ filter, and $K$-band magnitude determined with the KPNO $K$-band filter (Dickinson et~al. 2000; M.~Dickinson, 2000, private communication; Waddington et~al. 1999).}

\newpage

\end{deluxetable}

\clearpage

%
%%%%%%%%%%%%%%%%%%%%%%%%%%%%%%%%%%%%%%%%%%%%%%%%%%%%%%%%%%%%%%%%%%%%%%
% TABLE 2: x-ray detected uJy sources
%%%%%%%%%%%%%%%%%%%%%%%%%%%%%%%%%%%%%%%%%%%%%%%%%%%%%%%%%%%%%%%%%%%%%%
%

%
%%%%%%%%%%%%%%%%%%%%%%%%%%%%%%%%%%%%%%%%%%%%%%%%%%%%%%%%%%%%%%%%%%%%%%
% TABLE 2
%%%%%%%%%%%%%%%%%%%%%%%%%%%%%%%%%%%%%%%%%%%%%%%%%%%%%%%%%%%%%%%%%%%%%%
%

\begin{deluxetable}{lcccccllccccrrrl}
\rotate
\tablecolumns{16}
\tabletypesize{\tiny}
\tablewidth{0pt}
\tablecaption{X-ray detected optically faint $\mu$Jy radio sources}
\tablehead{
\colhead{} &
\colhead{} &
\colhead{} &
\colhead{Radio} &
\colhead{} &
\colhead{} &
\multicolumn{2}{c}{X-ray coordinates} &
\colhead{} &
\multicolumn{3}{c}{Counts}      &
\multicolumn{3}{c}{Flux}  &
\colhead{Further X-ray} \\
\colhead{Name$^{\rm a}$} &
\colhead{$S_{1.4GHz}$$^{\rm b}$}      &
\colhead{$\alpha$$^{\rm c}$}          &
\colhead{ID$^{\rm d}$}          &
\colhead{$I^{\rm e}$}                  &
\colhead{$I-K^{\rm f}$}                 &
\colhead{$\alpha_{2000}$}       &
\colhead{$\delta_{2000}$}       &
\colhead{$R-X$$^{\rm g}$} &
\colhead{FB$^{\rm h}$}          &
\colhead{SB$^{\rm h}$}          &
\colhead{HB$^{\rm h}$}          &
\colhead{FB$^{\rm i}$}          &
\colhead{SB$^{\rm i}$}          &
\colhead{HB$^{\rm i}$}          &
\colhead{References}}
\startdata
  VLA J123618+621550$^{\rm j}$&         $150\pm{8}$&             $>$0.63&         U&             $>$25.3&                 $-$&                        12 36 18.38&                        +62 15 51.1&            0.7&                       16.5$\pm$6.1&                            $<$13.2&                            $<$13.3&      0.19&   $<$0.06&   $<$0.31&                    \\
            VLA J123636+621320&          $50\pm{8}$&             $>$0.90&         U&             $>$25.3&              $>$3.4&                        12 36 36.89&                        +62 13 20.2&            0.2&                       18.2$\pm$6.3&                        7.6$\pm$4.4&                       11.9$\pm$5.6&      0.20&      0.04&      0.27&                    \\
            VLA J123642+621331&         $470\pm{6}$&                0.94&         U&      24.9$^{\rm k}$&       3.5$^{\rm k}$&                        12 36 42.11&                        +62 13 31.6&            0.2&                       22.7$\pm$6.8&                       20.4$\pm$6.0&                             $<$9.5&      0.25&      0.10&   $<$0.21&            Paper IV\\
            VLA J123646+621448&         $120\pm{9}$&                0.84&         U&                24.9&              $<$3.8&                        12 36 46.05&                        +62 14 49.0&            0.3&                       12.6$\pm$5.7&                        9.2$\pm$4.6&                             $<$8.8&      0.14&      0.04&   $<$0.20&                    \\
            VLA J123651+621221&          $49\pm{6}$&                0.71&         S&      25.8$^{\rm k}$&       3.7$^{\rm k}$&                        12 36 51.75&                        +62 12 21.4&            0.1&                     148.1$\pm$14.2&                       60.1$\pm$9.2&                      91.9$\pm$11.6&      2.81&      0.28&      2.59&  Papers I,II, \& IV\\
  VLA J123701+621146$^{\rm j}$&        $130\pm{10}$&                0.67&         S&                24.8&                5.6&                        12 37 01.66&                        +62 11 45.9&            0.9&                       11.8$\pm$4.4&                        7.2$\pm$2.8&                             $<$7.4&      0.14&      0.04&   $<$0.18&                    \\
            VLA J123707+621408&          $45\pm{6}$&                0.29&       AGN&                25.0&                 5.0&                        12 37 07.23&                        +62 14 08.0&            0.2&                       54.7$\pm$9.3&                       24.2$\pm$6.4&                       32.2$\pm$7.7&      1.00&      0.12&      0.92&                    \\
            VLA J123711+621325&          $54\pm{9}$&             $>$1.16&         S&                25.0&              $<$3.9&                        12 37 11.97&                        +62 13 25.2&            0.5&                       22.2$\pm$6.9&                             $<$9.7&                       20.4$\pm$6.7&      0.56&   $<$0.05&      0.72&            Paper II\\
  VLA J123721+621130$^{\rm j}$&         $380\pm{5}$&             $-$0.28&       AGN&             $>$25.3&                 $-$&                        12 37 21.18&                        +62 11 29.7&            0.6&                       21.5$\pm$6.4&                             $<$9.9&                       12.7$\pm$4.6&      0.29&   $<$0.06&      0.35&                    \\
\enddata

\tablenotetext{a}{Radio source name. Taken from Richards et~al. (1999).}
\tablenotetext{b}{Radio flux density at 1.4 GHz in units of $\mu$Jy. Taken from Richards et~al. (1999).}
\tablenotetext{c}{Radio spectral index, where $F_{\nu}\propto\nu^{-\alpha}$. Taken from Richards et~al. (1999).}
\tablenotetext{d}{Source classification based on radio properties. Taken from Richards (1999).}
\tablenotetext{e}{$I$-band Vega-based magnitude.}
\tablenotetext{f}{Calculated $I-K$ color determined from $I-(HK^\prime-0.3)$ following Barger et~al. (1999).}
\tablenotetext{g}{Positional offset between radio source and X-ray source in arcsecs.}
\tablenotetext{h}{Source counts and errors. Taken from Table~1 for sources detected with {\sc wavdetect} false probability threshold of 10$^{-7}$ and from {\sc wavdetect} for sources detected with {\sc wavdetect} false probability threshold of 10$^{-5}$. ``FB'' indicates full band, ``SB'' indicates soft band, and ``HB'' indicates hard band.}
\tablenotetext{i}{Fluxes are in units of $10^{-15}$~erg~cm$^{-2}$~s$^{-1}$. These fluxes have been calculated following the method described in Paper V. They have not been corrected for Galactic absorption. ``FB'' indicates full band, ``SB'' indicates soft band, and ``HB'' indicates hard band.}
\tablenotetext{j}{X-ray counterpart detected with {\sc wavdetect} with false probability threshold of 10$^{-5}$.}
\tablenotetext{k}{$I$-band magnitude determined with the WFPC2 $I814W$ filter, and $K$-band magnitude determined with the KPNO $K$-band filter (Dickinson et~al. 2000; M.~Dickinson, 2000, private communication; Waddington et~al. 1999).}

\newpage

\end{deluxetable}

\clearpage

%
%%%%%%%%%%%%%%%%%%%%%%%%%%%%%%%%%%%%%%%%%%%%%%%%%%%%%%%%%%%%%%%%%%%%%%
% TABLE 3: x-ray stacking results
%%%%%%%%%%%%%%%%%%%%%%%%%%%%%%%%%%%%%%%%%%%%%%%%%%%%%%%%%%%%%%%%%%%%%%
%

%%%%%%%%%%%%%%%%%%%%%%%%%%%%%%%%%%%%%%%%%%%%%%%%%%%%%%%%%%%%%%%%%%%%%%
% TABLE 3
%
% Results of 970 ks stacking analysis...
%%%%%%%%%%%%%%%%%%%%%%%%%%%%%%%%%%%%%%%%%%%%%%%%%%%%%%%%%%%%%%%%%%%%%%
%

\begin{deluxetable}{lcccccccccccrrr}
%\rotate
\tabletypesize{\tiny}
\tablewidth{0pt}
\tablecaption{X-ray stacking analysis results}

\tablehead{
\colhead{Sample}&
\colhead{} &
\colhead{Effective}&
\multicolumn{3}{c}{Source counts}&
\multicolumn{3}{c}{Background counts}&
\multicolumn{3}{c}{Statistical probability}&
\multicolumn{3}{c}{Flux}\\
\colhead{Reference$^{\rm a}$} &
\colhead{$N^{\rm b}$}& 
\colhead{Exposure$^{\rm c}$}& 
\colhead{FB$^{\rm d}$}& 
\colhead{HB$^{\rm d}$}&  
\colhead{SB$^{\rm d}$}&  
\colhead{FB$^{\rm e}$}& 
\colhead{HB$^{\rm e}$}&  
\colhead{SB$^{\rm e}$}&  
\colhead{FB$^{\rm f}$}& 
\colhead{HB$^{\rm f}$}&  
\colhead{SB$^{\rm f}$}&  
\colhead{FB$^{\rm g}$}& 
\colhead{HB$^{\rm g}$}&  
\colhead{SB$^{\rm g}$}
}
\startdata
Conti \etal (1999)         & 7  & 6.5 & 41  & 21 &  5 & 38.3 & 23.2 &  6.5 & 0.35 & 0.70 & 0.78 & $<$2.1 & $<$3.9 & $<$0.5 \\
Jarvis \& MacAlpine (1998) & 12 & 11.1 & 56  & 27 &  17 & 68.0 & 39.1 & 14.9 & 0.79 & 0.89 & 0.32 & $<$1.5 & $<$2.6 & $<$0.5 \\
Richards \etal (1999)      & 8 & 7.2 & 53 & 29 & 16 & 41.7 & 24.0 & 8.7 & $5.1\times10^{-2}$ & 0.18 & $1.7\times10^{-2}$ & $<$2.2 & $<$4.2 & 0.5 \\
\enddata
\tablenotetext{a}{Object sample reference.}
\tablenotetext{b}{Total number of sources used in the stacking analysis.}
\tablenotetext{c}{Effective \chandra\ exposure time in Ms.}
\tablenotetext{d}{Total counts measured. ``FB'' indicates full band, ``HB'' indicates hard band, and ``SB'' indicates soft band.}
\tablenotetext{e}{Calculated number of expected background counts. ``FB'' indicates full band, ``HB'' indicates hard band, and ``SB'' indicates soft band.}
\tablenotetext{f}{Poisson probability that the total number of counts measured could be due to statistical chance. ``FB'' indicates full band, ``HB'' indicates hard band, and ``SB'' indicates soft band.}
\tablenotetext{g}{Average flux or $3\sigma$ upper limit in units of $10^{-17}$~erg~cm$^{-2}$~s$^{-1}$. ``FB'' indicates full band, ``HB'' indicates hard band, and ``SB'' indicates soft band.}

\newpage

\end{deluxetable}

\clearpage

%\epsscale{0.7}

\renewcommand{\baselinestretch}{1.0}

%%%%%%%%%%%%%%%%%%%%%%%%%%%%%%%%%%%%%%%%%%%%%%%%%%%%%%%%%%%%%%%%%%%%%%

\end{document}